\documentclass[11pt]{article}

\usepackage{bm}
\usepackage{algorithmic}
\usepackage{algorithm}
\usepackage{latexsym}
\usepackage{amssymb}
\usepackage{amsbsy}
\usepackage{amsmath}
\usepackage{graphicx}
\usepackage{subfigure}
\usepackage{url}
\usepackage{enumitem}
\usepackage{multirow}

\newcommand{\idc}{\bm{1}}
\newcommand{\N}{\mathcal{N}}

\newcommand{\D}{\mathcal{D}}
\newcommand{\E}{\mathcal{E}}
\newcommand{\G}{\mathcal{G}}
\newcommand{\Ex}{\mathbb{E}}
\newcommand{\pr}{\mathbb{P}}
\newcommand{\p}{\mathbf{P}}
\newcommand{\z}{\mathbf{Z}}

\newcommand{\rt}{\right}
\newcommand{\lt}{\left}

\newcommand{\R}{\mathbb{R}}
\newcommand{\var}{\mathrm{Var}}
\newcommand{\cov}{\mathrm{Cov}}
\newcommand{\ssigma}{\boldsymbol\sigma}
\newcommand{\eeta}{\boldsymbol\eta}

\newtheorem{definition}{Definition}
\newtheorem{lemma}{Lemma}

\newtheorem{proposition}{Proposition}

\begin{document}

\title{Exploiting the Past to Reduce Delay in CSMA Scheduling: \\ A High-order Markov Chain Approach}

\author{\vspace{3mm}
Jaewook Kwak, Chul-Ho Lee, and Do Young Eun \\
Department of Electrical and Computer Engineering \\
North Carolina State University, Raleigh, NC 27695 \\
Email: \{jkwak, clee4, dyeun\}@ncsu.edu}

\date{February, 2013}
\maketitle

\pagestyle{plain} \setcounter{page}{1} \pagenumbering{arabic}

\renewcommand{\baselinestretch}{1.1} 

\begin{abstract}
Recently several CSMA algorithms based on the Glauber dynamics model have been proposed for multihop wireless scheduling, as viable solutions to achieve the throughput optimality, yet are simple to implement. However, their delay performances still remain unsatisfactory, mainly due to the nature of the underlying Markov chains that imposes a fundamental constraint on how the link state can evolve over time.
In this paper, we propose a new approach toward better queueing and delay performance, based on our observation that the algorithm needs not be Markovian, as long as it can be implemented in a distributed manner, achieve the same throughput optimality, while offering far better delay performance for general network topologies. Our approach hinges upon utilizing past state information observed by local link and then constructing a high-order Markov chain for the evolution of the feasible link schedules. We show in theory and simulation that our proposed algorithm, named \emph{delayed CSMA}, adds virtually no additional overhead onto the existing CSMA-based algorithms, achieves the throughput optimality under the usual choice of link weight as a function of local queue length, and also provides much better delay performance by effectively `de-correlating' the link state process (thus removing link starvation) under any arbitrary network topology. From our extensive simulations we observe that the delay under our algorithm can be often reduced by a factor of 20 over a wide range of scenarios, compared to the standard Glauber-dynamics-based CSMA algorithm.
\end{abstract}

\vspace{2mm}


\section{Introduction}

Medium access control (MAC) algorithm, which decides link level data transmission, is a central component in wireless packet scheduling. As the MAC plays an important role in achieving efficient channel utilization and providing quality of service for diverse wireless applications, designing an efficient MAC algorithm has been considered to be of significant importance. In a rich and long history of research on this subject, the most commonly believed goal is to achieve the following properties: (i) high-throughput utilization, (ii) low delay, and (iii) simple and distributed implementation. These three criteria, however, have different trade-offs among them, and hence developing an algorithm that has all the properties simultaneously is still a challenging problem.

In wireless networks, the property of high-throughput is often characterized by the packet arrival rate region under which the algorithm stabilizes the network queues. A classical algorithm, the max-weight scheduling (MWS) \cite{TE92}, is known to achieve the largest rate region under a general independent set constraint model. The MWS is however, not deemed practical since it requires global information to solve a complicated combinatorial optimization problem in each time instance. Many heuristics such as greedy-maximal scheduling (GMS) or maximal-matching algorithms are considered as alternatives to MWS, but they may achieve only a fraction of the capacity region~\cite{CapacityGMS08, LimitGMS07, EfficiencyGMS06, ImpactImperfect05}, or are throughput-optimal only on certain types of network topology~\cite{LocalPool06, LocalMax08, ImprovedBoundGMS09}. There are also several methodologies for achieving the throughput-optimality in general network~\cite{TOPartition06, TOGossip06, ConstantOverhead07}, but they have turned out to incur excessive message passing in many cases.

Recently, a great advance has been made toward this problem in a class of CSMA scheduling, where the throu-ghput-optimality can be achieved in a simple and distributed manner, e.g.~\cite{JW10DC,JLNSW11TIT,SRS09,Shah12-AAP,JRJ10CDC-techrep}. The idea is based on the so-called Glauber dynamics, which is a Monte Carlo Markov Chain (MCMC) method that often provides an approximate solution to a combinatorial optimization problem. The key enabler in realizing the method for the CSMA scheduling is achieving a desired probability distribution for the max-weight schedule by locally controlling the CSMA parameters without explicit knowledge of arrival rate or neighboring information. For example, in \cite{JLNSW11TIT}, an algorithm was developed to adaptively choose the parameter with local queue information, and the throughput-optimality is proved under the time-scale decomposition assumption (the system converges to its stationary regime quickly enough before adaptation of the CSMA parameters). In an another approach \cite{SRS09}, the optimality is established without the assumption by taking the parameters as slowly varying function of the queue-size. More general sufficient conditions on the function for the throughput-optimality has been studied in \cite{JRJ10CDC-techrep}.

Although the fact that these CSMA algorithms guarantee the throughput-optimality is an appealing merit, simulations often demonstrate that they incur large backlogs on the queue, and the resulting delay performance is far from being satisfactory. In the literature, there have been several attempts to improve the delay performance. For example, in~\cite{Shah10DO}, the authors adapt the graph-coloring scheme to the CSMA algorithm so as to achieve a constant-bounded property on the average delay. However, such an improvement of delay comes at the cost of sacrificing achievable rate region.
\cite{Marbach11TO} proposes an algorithm called U-CSMA that attempts to resolve link starvation problem by periodically resetting the ongoing schedules to an empty schedule. For torus and grid types of topologies, this idea provides an average delay bounded by a constant factor independent of the network size. However, such a good delay performance is not guaranteed in general network topologies.
\cite{Kklam12} shows that the delay performance can be improved when multiple channels are available while only one can be used for data transmission, but such multiple channels may not be present in general.
And, an idea in \cite{Dongyue10Allerton} prioritizes link scheduling by deferring links with shorter queues in order to reduce delays in longer queues. In this idea, however, the delay performance on the link with smaller queue can be even worse.
In \cite{Clee12}, the authors propose a generalized version of Glauber dynamics that all achieve the same stationary distribution and suggest that Metropolis-Hastings algorithm gives better delay performance.

In this paper, we propose a new approach toward better queueing and delay performance, based on our observation that the algorithm needs not be Markovian, as long as it can be implemented in a distributed manner, achieve the same throughput optimality, while offering far better delay performance for general network topologies. Our algorithm, termed \emph{delayed CSMA}, updates next schedule not based on the current status, but on several step-back past state information, thus necessitating high-order Markov chain modeling. This schedule update based on `delayed' information, somewhat counter-intuitively, provides a significant gain in delay performance by effectively removing the strong correlations that persist in the link state process and thus removing link starvation problems. We prove that our algorithm achieves the throughput optimality, yet provides much better delay performance in the steady state by `reshaping' the correlation structure to our advantage, while keeping the stationary distribution of the schedules intact. Our extensive simulations show that the delay under our algorithm is smaller than the conventional CSMA algorithm by often a factor of 20 over a wide range of scenarios. Our analysis also offers an interesting viewpoint about the role of the mixing time on the delay performance~\cite{JLNSW11TIT,SubramanianISIT11} by showing the tradeoff between faster mixing time in the transient phase and smaller correlations in the steady state.

The rest of this paper is organized as follows. In Section 2, we present our network model and the Glauber dynamics for CSMA algorithms, as well as some definitions on capacity region and the throughput optimality. In Section 3, we first explain our motivation and our own approach, and then show that there indeed exists strong correlations in the link state process (the service process of the queue at the link) under the standard Glauber dynamics based CSMA algorithm. We then construct a class of high-order Markov chains (of order $T$) and present its distributional and correlation properties in the steady-state. In Section 4, we first show how our algorithm leads to better delay performance, and then prove that our algorithm is also throughput optimal. And, we also discuss some tradeoff induced by our algorithm. Section 5 presents our extensive simulation results under various network scenarios, and we conclude in Section 6.


\section{Preliminaries}

\subsection{Network Model}

We consider a wireless network with a \emph{conflict graph} $\G = (\N, \E)$ where $\N$ is the set of links (transmitter-receiver pair), and $\E$ is the set of edges which represents conflict relationship between links. An edge $(i,j) \in \E$ exists between two links $i$ and $j$ if simultaneous use of the two leads to failure of communications.
We define a schedule by $\ssigma = (\sigma_v)_{v\in \N} \in \{0,1\}^{|\N|}$, which represents the set of transmitting links. A link $v$ (or node $v$ in the conflict graph $\G$) is active if it is included in the schedule, i.e., $\sigma_v = 1$, and is inactive if otherwise. A \emph{feasible} schedule is a set of links that can be active at the same time slot according to the conflict relationship $\E$. Thus, a feasible schedule $\ssigma$ should satisfy the independent set constraint i.e., $\sigma_i + \sigma_j \leq 1$ for all $(i,j) \in \E$. We denote by $\Omega$ the set of all feasible schedules.

\begin{figure}[ht!]
    \centering
    \vspace{-0mm}
    \hspace{-5mm}\includegraphics[width=2.7in]{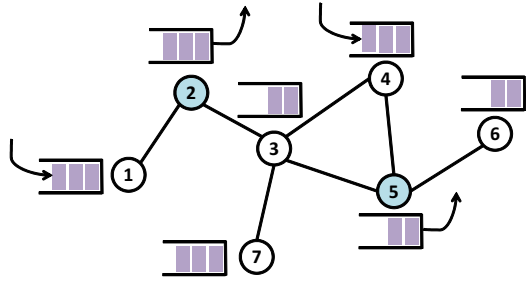}
    \vspace{-2mm}
    \caption{An instance of schedule where links 2 and 5 are active in a conflict graph with $|\N|=7$.} \label{fig:topology}
    \vspace{-2mm}
\end{figure}

In our model, each link is associated with a queue fed by some exogenous traffic arrivals and serviced when the link is active. We consider that a packet arrives to the queue of link $v$ at each time slot $t$ according to a Bernoulli process $A_v(t)$, i.e., $A_v(t)$, $t=1,2,\ldots$ are $i.i.d.$ with $\Ex\{A_v(t)\} = \eta_v$. Let $\eeta = (\eta_v)_{v \in \N}$ be the set of arrival rates to the queues in the network. Let $Q(t) = (Q_v(t))_{v \in \N}$ be the number of packets in the queue at time $t$. Then the queue dynamics is governed by the following recursion:
\begin{equation}
Q_v(t) = [Q_v(t\!-\!1) + A_v(t) - \sigma_v(t)]^+, ~~t\geq 1,
\label{eq:queue}
\end{equation}
where $[x]^+ = \max\{0,x\}$. See Figure~\ref{fig:topology} for illustration.

\subsection{CSMA Scheduling as a Glauber Dynamics}
\label{subse:CSMA}

The basic idea of throughput-optimal CSMA is to utilize the Glauber dynamics as a link scheduling algorithm.
Direct adaptation of the traditional Glauber dynamics to scheduling is as follows.
For a graph $\G$, at every time $t \in \mathbb{N}$, a link $v$ is chosen uniformly at random from $\N$. Then,
\begin{align*}
&\mbox{If}~\sum_{w \in N_v} \sigma_w(t-1) = 0, ~\mbox{then}~
\begin{cases} \sigma_v(t) = 1, &\mbox{w.p.}~\frac{\lambda_v}{1+\lambda_v}, \\
  \sigma_v(t) = 0, &\mbox{w.p.}~\frac{1}{1+\lambda_v},
\end{cases} \\
& \mbox{otherwise},~ \sigma_v(t)=0, \\
& \mbox{and for all}~ w\neq v, ~\mbox{set}~ \sigma_w(t) = \sigma_w(t-1),
\end{align*}
where $N_v = \{w: (v,w) \in \E\}$ is a set of neighboring nodes of $v$, and $\lambda_v$ is a parameter called \emph{fugacity}. Given $\lambda_v > 0$ for all $v \in \N$, the schedule $\ssigma(t)$ forms a Markov chain which is irreducible, aperiodic, and reversible over $\Omega$, and achieves the stationary distribution given by
\begin{equation}
\pi(\ssigma) = \frac{1}{Z}\prod_{i \in \N} \lambda_i^{\sigma_i}, \label{eq:glauber_stationary}
\end{equation}
where $Z = \sum_{\ssigma \in \Omega} \prod_{i \in \N} \lambda_i^{\sigma_i}$ is a normalizing constant.

A practical CSMA algorithm uses a modified version of the above procedure. First, the fugacity parameter $\lambda_v$ is set to be changing dynamically over time $t$. It is typically chosen as a function of locally observable information such as $Q_v(t)$. Second, multiple links are allowed to be selected in a single time slot~\cite{NTS10QCSMA,JLNSW11TIT}. In this procedure, a set of links that do not conflict with each other is selected. This can be achieved in a distributed fashion through a simple randomized procedure. For instance, each link attempts to access the channel with access probability $a_v$, $v\in \N$, and link $v$ is then selected with probability
\begin{equation}
m_v = a_v \prod_{j \in N_v} (1-a_j).
\label{eq:mv}
\end{equation}
The set of chosen links $D(t)$ following this procedure is called a  \emph{decision schedule} at time $t$.
More practical implementation tailored to IEEE 802.11 can be found in~\cite{NTS10QCSMA}.

It is not difficult to find a continuous time version of the above model, however as noted in \cite{NTS10QCSMA, AccessProb12Infocom}, the discrete time model has several advantages over a continuous one. For example, by synchronizing the time slots, the well known hidden and exposed terminal problems can be effectively resolved. And also, continuous time model often requires perfect channel sensing mechanism which may not be feasible in practice. In discrete case, the time needed for deciding decision schedules can be bounded by a constant fraction, and hence the maximum throughput can be established. For these reasons, we assume the discrete-time CSMA algorithm throughout the paper.

\subsection{Capacity Region and Stability} \label{subse:TO-prelim}

The capacity region of the network is the set of all arrival rates $\eeta$ for which there exists a scheduling algorithm that can support the arrivals while preserving network queues stable. It is known \cite{TE92} that the capacity region is given by the convex combination of all feasible schedules, i.e.,
\begin{equation*}
\mathbb{C} = \left\{ \sum_{\ssigma \in \Omega} \theta_{\ssigma}\ssigma : \sum_{\ssigma \in \Omega} \theta_{\ssigma} = 1, \theta_{\ssigma} \geq 0, \forall \ssigma \in \Omega \right\}
\end{equation*}

We here first collect several definitions. Let $W_v(t)$ be a weight function associated with a link $v \in \N$ at time slot $t$. It was shown in \cite{TE92} that MWS is throughput-optimal with $W_v(t) = Q_v(t)$ (see below for definition in more general settings), provided that it can select a maximum-weight schedule $\ssigma^*(t)$ in every time slot where
\[
\ssigma^*(t) = \arg\max_{\ssigma \in \Omega}  \sum_{v \in \ssigma} W_v(t).
\]
This has been generalized in \cite{ESP05} as follows. For all $v \in \N$, set link weights as $W_v(t) = h(Q_v(t))$ for some monotone increasing functions $h:[0,\infty) \rightarrow [0, \infty)$. (See \cite{ESP05} for precise definitions for the weight functions $h(\cdot)$.) In this case, a scheduling algorithm is said to be \emph{throughput optimal} if for all arrival rates inside the capacity region, network queues are stable in the sense that
\begin{equation}
\limsup_{K \rightarrow \infty} \frac{1}{K} \sum_{t=0}^{K-1} \Ex \left[ \left( \sum_{v \in \N} h^2(Q_v(t)) \right)^{1/2} \right] < \infty.
\label{eq:stability}
\end{equation}

Recently, the authors in \cite{SRS09,Shah12-AAP} have shown that with the choice of $\lambda_v(t) = e^{W_v(t)}$ where the weights  $W_v(t)$ are in the form of $\log\log(Q_v(t))$, the conventional CSMA algorithm via Glauber dynamics as in Section~\ref{subse:CSMA} achieves the throughput optimality. In~\cite{JRJ10CDC-techrep}, the choice of $\log\log(\cdot)$ for $h(\cdot)$ has been slightly generalized into $\log(\cdot)/g(\cdot)$ for a function $g$ that increases arbitrarily slowly. When the whole system including all the queue-lengths is a Markov chain, the condition in (\ref{eq:stability}) implies that the chain is positive recurrent~\cite{Meyn92,JRJ10CDC-techrep,Shah12-AAP}. It is however worth mentioning that the condition in (\ref{eq:stability}) itself is established under a fairly general case in that the system of $Q_v(t)$ doesn't need to be a Markov chain.

\section{CSMA Scheduling with Delayed Glauber Dynamics}

\subsection{Motivation and Our Approach}

According to the standard queueing theory, the queueing delay is governed by not only the long-term average arrival and service rates, but also their higher-order statistics such as their correlations or dependency over time~\cite{BigQueues,Bremaud03}. Indeed, there are a number of works in the literature suggesting that positive correlations have an adverse impact on the queueing delay~\cite{Addie94a,Duffield94a,Duffield95a,Kelly96}. With this in mind, in this paper, we aim at developing a new, distributed algorithm similar to the current CSMA-based ones~\cite{SRS09,JRJ10CDC-techrep,Shah10DO,NTS10QCSMA,Shah12-AAP,Clee12}, but offering far superior delay performance by effectively reducing such correlations in any general network topology, while keeping the throughput optimality intact.

\begin{figure}[t]
    \centering
    \subfigure[Conventional CSMA algorithm]{ \includegraphics[width=2.5in]{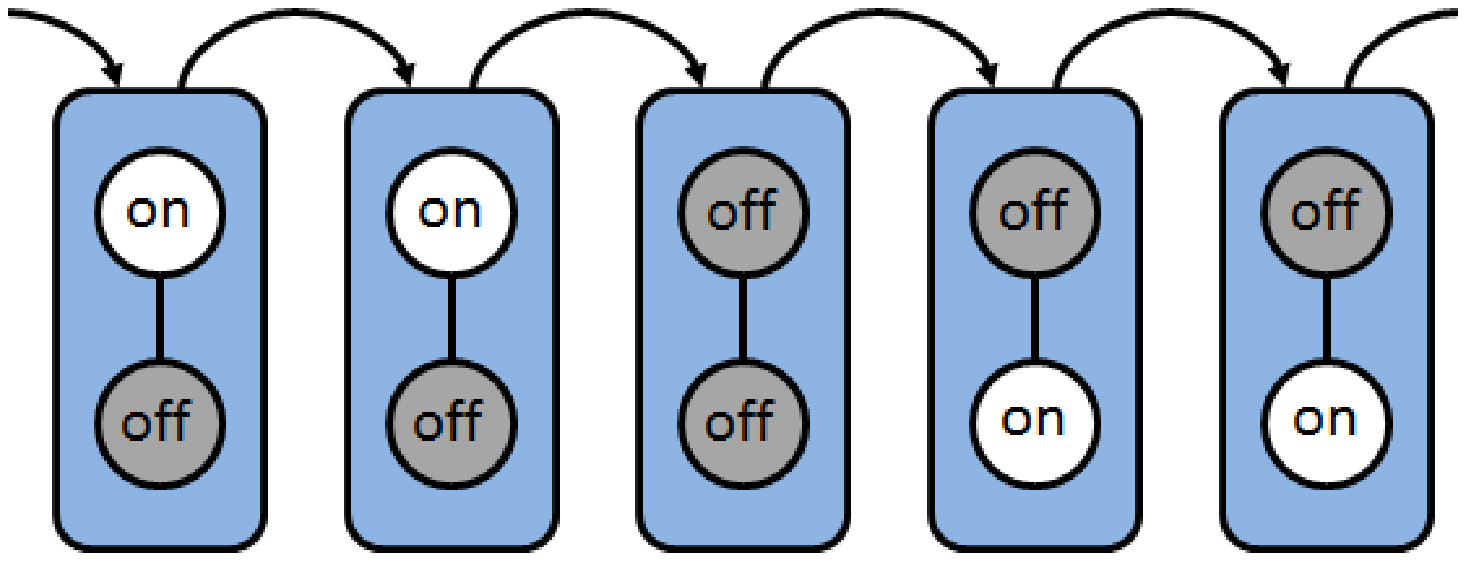} \label{fig:tr_figure1} }
    \hspace{10mm}
    \subfigure[Proposed approach: Delayed CSMA]{  \includegraphics[width=2.5in]{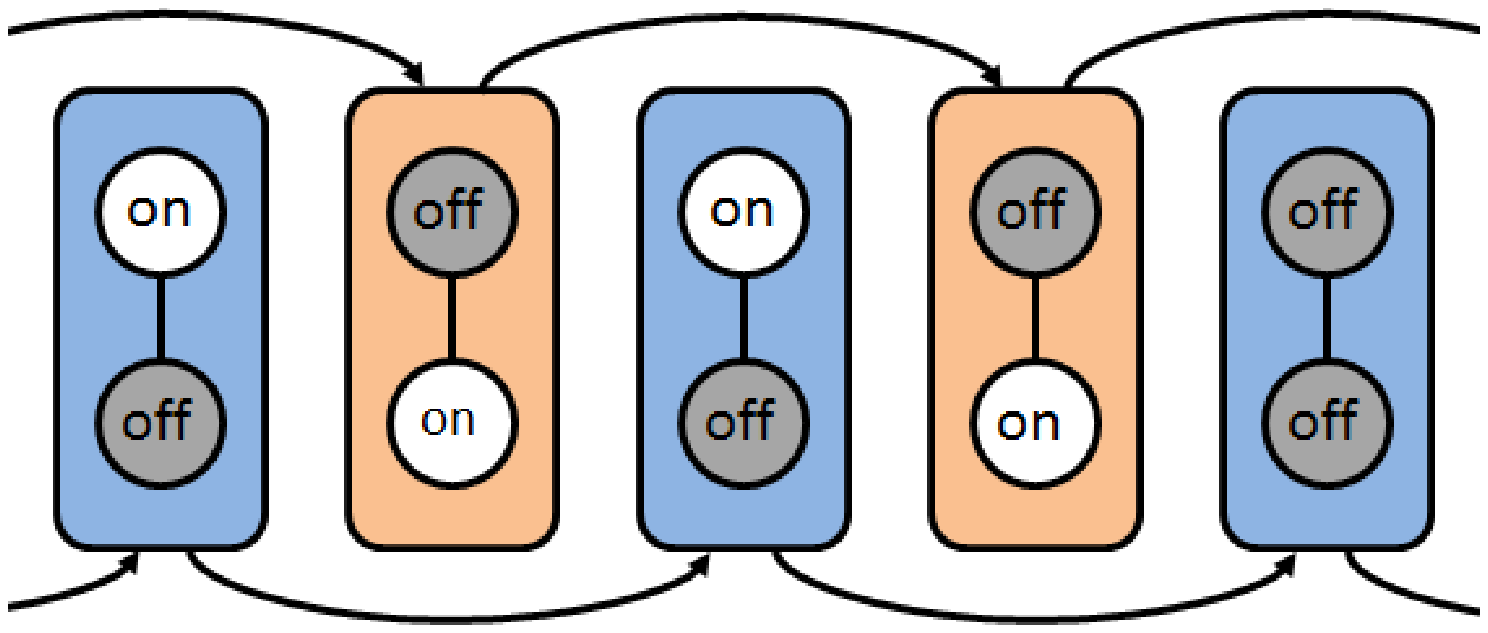} \label{fig:tr_figure2}}
    \caption{Comparison between the conventional CSMA and our proposed approach. A box indicates a schedule, and arrows indicate state transitions.} \label{fig:tr_figure}
    \vspace{-2mm}
\end{figure}

Our motivation comes from the fact that the service process $\sigma_v(t)$ at link $v$ under the standard CSMA policy is often heavily correlated over time. (We will discuss it in more detail in Section~\ref{subsection:correlation}.) This is because once CSMA finds a schedule, it tends to stay in the same schedule or its similar set of schedules for a long time~\cite{Marbach11TO}.
To illustrate, consider for example two links that are close to each other, so that only one link can be active at a time. At a particular moment, if a schedule of the two links is `active-inactive', the inactive link first has to wait until the active link release the channel occupation. In this case, transition to next possible set of states is limited to `active-inactive', and then `inactive-inactive states'. Direct transition to `inactive-active' state is impossible. (See Figure~\ref{fig:tr_figure1} for example.) This phenomenon hinders a frequent switch between schedules, leading to starvation for the corresponding inactive link.

The method we propose in this paper effectively resolves this problem. The main idea is as follows. Suppose we have two schedulers that respectively generate schedules independently, while preserving the feasibility constraint for each time slot. If we choose to use one scheduler at every odd time index, and the other one at every even time index (see Figure~\ref{fig:tr_figure2}), it is now possible to make a transition from `active-inactive' directly to `inactive-active' state, which would be impossible under the conventional CSMA. This alternate use of different schedulers produces more drastic change of states in consecutive time slots, thereby alleviating link starvation while maintaining the same long-term frequency of being active.

Given the potential benefits, we generalize this idea to the use of multiple schedulers in a round-robin manner. Throughout the paper, we will use a notation $T$ to indicate the number of such schedulers. In practice, this can be easily implemented in a distributed setting by having all links together update their schedules based on $T$-step-back state. For this purpose, each link only needs to remember its last $T$ channel states. This way, the whole system behaves \emph{as if} there are $T$ separate schedulers (or chains) taking turns to generate next schedules. Building upon this idea and applying to the CSMA scheduling, we next present our proposed algorithm, named \emph{delayed CSMA}. Note that if $T=1$, our algorithm reduces to the conventional CSMA-based scheduling algorithm.

\begin{algorithm}[h] \label{algol:XYZ_CSMA}
\caption{Delayed CSMA}
\begin{algorithmic}[1]
 \STATE Initialize: for all links $i \in \N$, $\sigma_i(t) = 0$, $t = 0,1,\ldots,T\!-\!1$.
 \STATE At each time $t\geq T$: links find a decision schedule, $\D(t)$ through a randomized procedure, and
 \FORALL{links $i \in \D(t)$}
   \IF{$\sum_{j \in N_i} \sigma_j(t-T) = 0$}
     \STATE $\sigma_i(t) = 1$ with probability $\frac{\lambda_i}{1+\lambda_i}$
     \STATE $\sigma_i(t) = 0$ with probability $\frac{1}{1+\lambda_i}$
   \ELSE
     \STATE $\sigma_i(t) = 0$
   \ENDIF
 \ENDFOR
 \FORALL{links $j \notin \D(t)$}
   \STATE $\sigma_i(t) = \sigma_i(t-T)$
 \ENDFOR
\end{algorithmic}
\end{algorithm}

\subsection{Understanding Correlation Structure of the Standard CSMA} \label{subsection:correlation}

Before proceed to explain the details about our algorithm and its benefit in reducing correlations, in this section, we first study how much correlations are present in the service process $\sigma_v(t)$ for the queue at each link, induced by the standard Glauber dynamics.

To set the stage, consider a homogeneous Markov chain $\ssigma(t) \in \Omega$ denoting a feasible configuration by the Glauber dynamics at time slot $t$, assuming fugacity parameter $\lambda_v$ is set to be a constant. Since we are interested in the long-run behavior of queueing performance, without loss of generality, we can assume that the Markov chain $\ssigma(t)$ is in its stationary regime, i.e., $\pr\{\ssigma(t) = \ssigma\} = \pi(\ssigma)$, for all $t \geq 0$.

We write $\pi(B_v) \triangleq \sum_{\ssigma \in B_v} \pi(\ssigma) = \pr_{\pi}\{\sigma_v(t)=1\}$ for the long-term proportion of service availability at link $v$, where $B_v \triangleq \{\ssigma \in \Omega : \sigma_v = 1\} \subseteq \Omega$ is a set of all feasible schedules for which $v$ is active. For the rest of the paper, we will mostly focus on the service process and queueing dynamics at a given link $v\in\N$, so we will drop the subscript $v$ and write $\sigma(t)$ for $\sigma_v(t)$ (similarly $\eta$ for $\eta_v$, $A(t)$ for $A_v(t)$, and $Q(t)$ for $Q_v(t)$) unless otherwise necessary.

We first state the following lemma which is useful in understanding the correlation structure of the service process $\sigma(t)$ under Glauber dynamics. An analogous statement for continuous time version was given by Lemma~3.6 of \cite{Sinclair07}, and we reproduce it here for a discrete-time case.

\begin{lemma}\cite{Aldous,Sinclair07}
Let $X_k$ be an irreducible, aperiodic, and reversible Markov chain with its transition probability matrix $P$ and stationary distribution $\pi$. If $X_0$ is drawn from $\pi$, then for $B \subset \Omega$, there exist $\alpha_j \geq 0$, $j=1,\ldots,|\Omega|$ such that
\begin{equation}
\pr\{X_k \in B | X_0 \in B\} = \sum_{j=1}^{|\Omega|} \alpha_j \rho_j^k, \label{eq:sinclair}
\end{equation}
where $\sum_j \alpha_j = 1$, $\alpha_1 = \pi(B)$, and $1=\rho_1 > \rho_2 \geq \cdots \geq \rho_{|\Omega|} > -1$  are eigenvalues of $P$.
\end{lemma}

Let $\psi(k) =\cov\{\sigma(t),\sigma(t+k)\}/\var\{\sigma(t)\}$ be the correlation coefficient of lag $k$ for the service process. Then, we have the following.

\begin{proposition} \label{lem:spectral}
Let $q_v = \sum_{\ssigma:\sigma_j = 0,\forall j \in N_v} \pi(\ssigma)$ be the probability that none of neighboring links of $v$ is active. Then,
\begin{align}
\psi(1) &= 1 - \frac{m_v}{1+(1-q_v)\lambda_v},  \label{lem:spectral-1} \\
\psi(2k) &\geq \left( 1 - \frac{m_v(2-m_v)}{1+(1-q_v)\lambda_v} \right)^k, ~~k=1,2,\ldots. \label{lem:spectral-2}
\end{align}
where $m_v$ is the probability that link $v$ is selected in a decision schedule as in (\ref{eq:mv}).
\end{proposition}

\vspace{2mm} \textbf{Proof:}
Let $E_t = \{\ssigma \in B_v\} = \{\sigma_v(t) = 1\}$ be the event that link $v$ is scheduled (active) at time $t$. Clearly, $\pr\{E_t\} = \Ex\{\sigma_v(t)\} = \pi(B_v)$. First, observe that
\begin{align*}
\cov\{\sigma(t), \sigma(t+k)\} & = \pr\{E_t, E_{t+k}\} - \pi(B_v)^2 = \pi(B_v)\lt(\pr\{E_{t+k} | E_t\} - \pi(B_v)\rt),
\end{align*}
and $\var\{\sigma(t)\} = \pi(B_v)(1 - \pi(B_v))$. Under $E_t$, link $v$ is active, thus all its neighbors must be inactive. In this situation, there are two possible events for link $v$ to stay active in the next time slot: (a) link $v$ is selected for update, but keep active state, (b) link $v$ is not selected.
The probability of event (a) is $\frac{m_v \lambda_v}{1+\lambda_v}$, and that of event (b) is $1-m_v$.
Thus,
\begin{equation*}
\pr\{E_{t+1} | E_t\} = \frac{m_v \lambda_v}{1+\lambda_v} + (1-m_v) =    1 - \frac{m_v}{1+\lambda_v}.
\end{equation*}
In \cite{JLNSW11TIT}, it was shown that $\pi(B_v) = \frac{\lambda_v}{1+\lambda_v} q_v$. (See Appendix B of \cite{JLNSW11TIT}.) Rearranging the terms, (\ref{lem:spectral-1}) follows.

Now consider the probability that if a link $v$ is active, it is also active after two time slots, i.e., $\pr\{E_{t+2} | E_t\}$. Let $E^c_t = \{\ssigma \notin B_v\}$ denotes an event that link $v$ is inactive at time $t$. Then,
\begin{align}
& \pr\{E_{t+2} | E_t\}  = \pr\{E_{t+2}, E_{t+1} | E_t\} + \pr\{E_{t+2}, E^c_{t+1} | E_t\} \nonumber
\end{align}
A simple calculation reveals that
$\pr\{E_{t+2}, E_{t+1}|E_t\} = (1-\frac{m_v}{1+\lambda_v})^2$, which is the probability that the link remains active for the next two consecutive time slots given that it is active now.

Similarly, $\pr\{E_{t+2}, E^c_{t+1} | E_t\}$ is the probability that the link $v$, from active state, changes to inactive and then active again. The first transition occurs with probability $\frac{m_v}{1+\lambda_v}$, and at this time note that none of the neighbors of $v$ is in the decision schedule. Thus, the second transition will occur with probability $\frac{m_v\lambda_v}{1+\lambda_v}$.
Thus, $\pr\{E_{t+2} | E_t\} = 1- \frac{m_v(2-m_v)}{1+\lambda_v}$, and we can write
\begin{equation*}
1 - \frac{m_v(2-m_v)}{1+\lambda_v} - \pi(B_v)  = \pr\{E_{t+2} | E_t\} - \pr\{E_t\} = \sum_{j=2}^{|\Omega|} \alpha_j \rho_j^2
\end{equation*}
where the second equality is from Lemma \ref{lem:spectral} and by choosing $B = B_v$ and $P$ as the transition probability matrix of the standard Glauber dynamics (Algorithm 1 with $T=1$). Note that
\begin{equation*}
\pr\{E_{t+2k} | E_t\} - \pr\{E_t\} = \sum_{j=2}^{|\Omega|} \alpha_j \rho_j^{2k}  = (1-\alpha_1)\frac{1}{1-\alpha_1} \sum_{j=2}^{|\Omega|} \alpha_j \rho_j^{2k}.
\end{equation*}
Define a random variable $Y\geq 0$ which takes value $\rho^{2}_j$ with probability $\frac{\alpha_j}{1-\alpha_1}$, $j=2,3,\ldots,|\Omega|$. Then the above can be written as $(1-\alpha_1) \Ex\{Y^k\}$. From Jensen's inequality, we have
\begin{equation*}
(1-\alpha_1) \Ex\{Y^k\} \geq (1-\alpha_1) \Ex\{Y\}^k
\end{equation*}
where the RHS is equal to $(1-\alpha_1) \left( 1 - \frac{m_v(2-m_v)}{(1-\alpha_1)(1+\lambda_v)}\right)^k$.
This proves (\ref{lem:spectral-2}) by noting that $\alpha_1 = \pi(B_v) = \frac{\lambda_v}{1+\lambda_v} q_v$.
\hfill $\Box$

\vspace{2.5mm}

Note that the obtained lower bounds are all positive since $0\leq m_v\leq 1$. Informed readers can also recognize that the lower bound can also be derived for all lags if the chain here were lazy, i.e., the transition matrix is $\frac{1}{2}(I+P)$ instead of $P$, so that all the eigenvalues are non-negative. Due to the possibility of eigenvalues being negative, the lower bound is, in general, valid only for even lags. In \cite{PracticalMCMC}, however, it is shown that for any finite state reversible Markov chain, the sum of adjacent pairs of correlations, $\psi(2n)+\psi(2n\!+\!1)$, is positive, decreasing, and convex in $n \geq 0$. Clearly, the standard Glauber dynamics is a reversible Markov chain on $\Omega$. This implies that, in conjunction with $\psi(1)$ being positive, the negative correlations in $\sigma(t)$, even if they exist, will be of much less magnitude and have lesser impact than positive ones.

\begin{figure}[t!]
    \centering
    \includegraphics[width=3.3in]{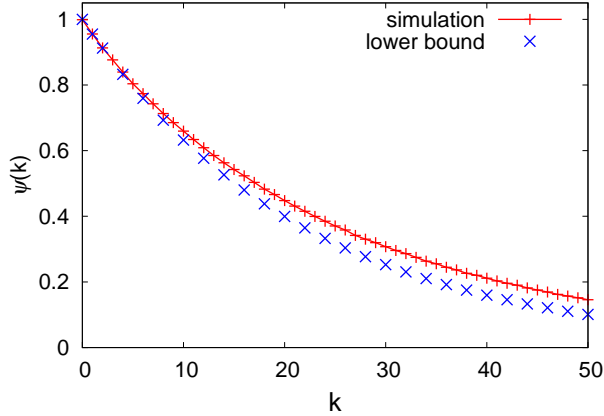}
    \vspace{-2mm}
    \caption{Correlations of $\sigma_3(t)$ for the network in Figure~\ref{fig:topology} from simulations and the lower bounds.} \label{fig:correlation}
    \vspace{-1mm}
\end{figure}

For instance, Figure~\ref{fig:correlation} shows measured correlations $\psi(k)$ for link $v=3$ in the network topology given in Figure~\ref{fig:topology}, based on the standard Glauber dynamics with access probability $a_v = 0.25$ and fugacity $\lambda_v = 1$ for all $v$, along with the predicted lower bounds in (\ref{lem:spectral-1}) and (\ref{lem:spectral-2}). We observe that the degree of correlations in the service process is significant and the lower bound in Lemma~\ref{lem:spectral} is in fact quite tight over a wide range of lags.

\subsection{High-order Markov Chain Model}\label{subse:high-order}

The key feature of our algorithm over the conventional one is that a new schedule is generated not from the current schedule but from $T$-step-back past schedule, and in doing so, each link just needs to make a decision based on its own $T$-step-back channel state information. Since every link operates this way, the independent set constraint is satisfied all the time.
From an analytical point of view, however, this means that the evolution of schedule $\ssigma(t)$ is no longer a Markov chain on $\Omega$, rendering all the results associated with Markov chains inapplicable.

Notwithstanding being non-Markov, our algorithm can still be modeled by a high-order Markov chain of order $T$ on the same state space $\Omega$, as follows:
\begin{align*}
& \pr\{X_t \!=\! x ~|~ X_{t\!-\!1} \!=\! x_{t\!-\!1}, \ldots, X_0 \!=\! x_0\} = \pr\{X_t \!=\! x ~|~ X_{t\!-\!1} \!=\! x_{t\!-\!1}, \ldots, X_{t\!-\!T} \!=\! x_{t\!-\!T}\},
\end{align*}
implying the current state depends upon $T$ past history.\footnote{Alternatively, one can also augment the state space into a product space $\Omega\times \cdots \times \Omega$ ($T$ times) on which $\{X_{t\!-\!T\!+\!1},\ldots, X_{t\!-\!1}, X_t\}$ becomes a Markov chain. But, this would lead to largely intractable descriptions and keep our exposition cluttered.} Our algorithm can then be succinctly written as
\begin{equation}
\pr\{X_t = y ~|~ X_{t\!-\!T} = x\} = P(x,y) \label{eq:high_order}
\end{equation}
where $P(x,y)$, $x,y \in \Omega$ is the transition probability of the conventional Markov chain from state $x$ to $y$.

We here collect several notations and simple facts that hold for finite state Markov chains~\cite{Aldous,Levin09,Bremaud99} and will prove useful throughout the analysis. Consider a finite-state, ergodic Markov chain $Y_n$ with its transition matrix $P$.
For functions $f, g:\Omega \rightarrow \R$, we define
\begin{equation*}
\langle f,\bold{P}^{(k)} g \rangle_{\mu} \triangleq \sum_{x,y \in \Omega}f(x)g(y)\mu(x)P^{(k)}(x,y)  = \Ex_{\mu}\{f(Y_0)g(Y_{k})\},
\end{equation*}
where $\mu$ is a probability distribution of $Y_0$ on $\Omega$, and $P^{(k)}(x,y)$ is $k$-step transition probability of the chain $Y_n$. For simplicity, we write $\langle f\rangle_{\mu} \triangleq \langle f, 1 \rangle_{\mu} = \Ex_{\mu}\{f(Y_0)\}$.
Also define
\begin{equation}
(\p^{(k)} f)(x)  \triangleq \sum_{y \in \Omega} P^{(k)}\!(x\!,y) f(x) = \Ex\{f(Y_k)|Y_0 \!=\! x\}.
\label{operator-P}
\end{equation}
For a (high-order) Markov chain $X_t$ with order $T$ as given in (\ref{eq:high_order}), if we define $Y^{m}_n = X_{nT + m}$ for $0 \leq m \leq T\!-\!1$ and $n = 0,1,2,\ldots$, then $\{Y^{m}_n\}_{n\geq 0}$ for each $m$ is a conventional Markov chain with initial state $Y^{m}_0 = X_m$. Since the chain $P$ is ergodic, we have, for any initial state $x$,
\begin{equation}
\lim_{k \rightarrow \infty}(\p^{(k)} f)(x) = \lim_{k \rightarrow \infty} \Ex\{f(Y^{m}_{k})|Y^{m}_0 = x\} = \langle f\rangle_{\pi}
\label{asdt}
\end{equation}
where $\pi$ is the stationary distribution of the chain $P$. Since this holds for any given $m=0,1,\ldots,T-1$, it follows that
\begin{equation*}
\lim_{t \rightarrow \infty} \Ex\{f(X_t)|X_0 = x_1, \ldots X_{T\!-\!1} = x_{T\!-\!1}\} = \langle f\rangle_{\pi},
\end{equation*}
implying that the marginal distribution of $X_t$ in the steady-state remains the same and does not change with $T$. However, we show next that different $T$ leads to strikingly different behavior in the  second order statistics.

\begin{proposition} (Asymptotic zero-correlation padding) \label{lm:zero_padding}
Let $X_t$ be a high-order Markov chain of order $T$ where the transition kernel is given by (\ref{eq:high_order}).
For any initial distribution, if $k \neq jT$, $j\in \mathbb{N}$,
\begin{equation*}
\lim_{t \rightarrow \infty} \Ex\{f(X_t)g(X_{t+k})\} = \langle f\rangle_{\pi} \langle g\rangle_{\pi},
\end{equation*}
assuming that the expectations exist.
\end{proposition}
\vspace{2mm} \textbf{Proof:}
Writing $t = nT+m$, and $t+k = n'T+m'$ for $m, m'\in \{0,\ldots,T\!-\!1\}$ and $n,n' = 0,1,\ldots$, one can verify that $m \neq m'$ if $k \neq jT$. Define $Y^{m}_{n} = X_{nT+m}$, and $Y^{m'}_{n'} = X_{n'T+m'}$. For given $m,m'$, $Y^{m}_{n}$ and $Y^{m'}_{n'}$ are both conventional Markov chains with transition kernel $P$. Then by conditioning, we have
\begin{align*}
\Ex\{f(X_t)g(X_{t+k})\} &= \Ex\{f(Y^{m}_{n})g(Y^{m'}_{n'})\} = \Ex\{ \Ex\{f(Y^{m}_{n})g(Y^{m'}_{n'}) | Y^m_0, Y^{m'}_{0}\}\},
\end{align*}
and observe that
\begin{align*}
\Ex\{f(Y^{m}_{n})g(Y^{m'}_{n'}) | Y^m_0 \!= i, Y^{m'}_{0} \! = j\} & = \Ex\{f(Y^{m}_{n}) | Y^m_0 \!= i, Y^{m'}_{0} \!= j\} \Ex\{g(Y^{m'}_{n'}) | Y^m_0 \!= i, Y^{m'}_{0} \!= j\} \\
& ~= \Ex\{f(Y^{m}_{n}) | Y^m_0 \!= i\} \Ex\{g(Y^{m'}_{n'}) | Y^{m'}_{0} \!= j\} \\
& ~ = (\p^{(n)} f)(i)(\p^{(n')} g)(j)
\end{align*}
where the first and second equalities follow from the conditional independence of $Y^{m}_{n}$ and $Y^{m'}_{n'}$ when initial values are fixed. Since $n,n' \to \infty$ as $t \to \infty$, by taking limits and from (\ref{asdt}), we are done.
\hfill $\Box$
\vspace{3mm}

\begin{proposition} (Asymptotic correlation-lag shifting) \label{lm:lag_shifting}
Let $X_t$ be a Markov chain of order $T$ with its transition kernel given by (\ref{eq:high_order}).
For any initial distribution and for any given $k \in \mathbb{N}$, we have
\begin{equation*}
\lim_{t \to \infty} \Ex\lt\{f(X_t)g(X_{t+kT})\rt\} = \langle f, \p^{(k)} g \rangle_{\pi},
\end{equation*}
assuming that the expectations exist.
\end{proposition}
\vspace{2mm} \textbf{Proof:}
As before, write $t = nT+m$ for $m \in \{0,1,\ldots,T\!-\!1\}$ and $n = 0,1,\ldots$. Then $Y^{m}_{n} = X_{nT+m}$ for each $m$ is a Markov chain. Observe that
\begin{equation*}
\Ex\{f(X_t)g(X_{t+kT})\} = \Ex\{f(Y^m_n)g(Y^m_{n+k})\}
= \langle f,\bold{P}^{(k)} g \rangle_{\mu_n^m},
\end{equation*}
where $\mu_n^m$ is the distribution of $Y^m_n$.
Since $n \rightarrow \infty$ as $t \rightarrow \infty$,
taking limit gives
\begin{equation*}
\lim_{n\to\infty} \langle f,\bold{P}^{(k)} g \rangle_{\mu_n^m} =  \langle f, \p^{(k)}g \rangle_{\pi},
\end{equation*}
since $\mu_n^m \Rightarrow \pi$ as $n \to \infty$ and the state space is finite. This completes the proof.
\hfill $\Box$
\vspace{3mm}

\begin{figure}[t!]
    \centering
    \vspace{-0mm}
    \hspace{-10mm}\includegraphics[width=3.3in]{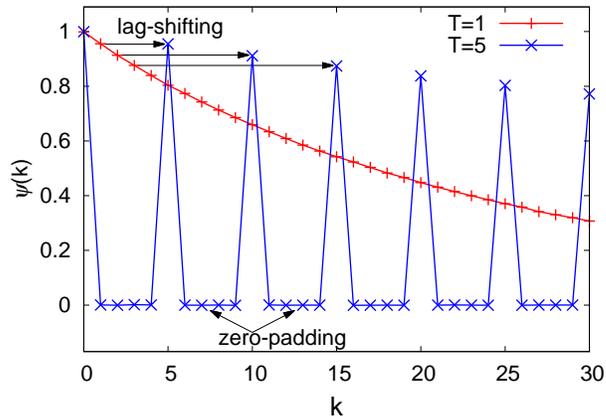}
    \vspace{-2mm}
    \caption{Two properties of the algorithm: correlation-lag shifting and zero-correlation padding} \label{fig:correlation-lemma}
    \vspace{-2mm}
\end{figure}

Recall that $\sigma_v(t) = \idc\{\ssigma(t) \in B_v\}$. Since $\ssigma(t)$ under our algorithm is a Markov chain of order $T$ (i.e., modeled as $X_t$ here), Propositions~\ref{lm:zero_padding} and \ref{lm:lag_shifting} tell us that the correlations $\psi(k)$ under our delayed CSMA algorithm with order $T$ are first shifted to $T$ times larger lags, and then all padded with zero in-between. To see this effect numerically, we have run the simulations with the same step as in Figure~\ref{fig:correlation}, but now with $T=5$. As seen in Figure~\ref{fig:correlation-lemma},
the correlations $\psi(1)$, $\psi(2)$, and $\psi(3)$ for $T=1$ case respectively gets shifted to $\psi(5)$, $\psi(10)$, and $\psi(15)$ for $T=5$, and $\psi(k)=0$ for all $k \not\equiv 0 \pmod{5}$. For the rest of the paper, we call these phenomena \emph{zero-padding} and \emph{lag-shifting}, respectively.

\section{Delay and Throughput Analysis}\label{se:analysis}

\subsection{Delay Performance}\label{subse:delay}

In this section, we investigate the efficiency of our delayed-CSMA algorithm of order $T$ in terms of its delay performance. As discussed earlier, our algorithm provides different second-order behavior while preserving the same long-term average statistics. Our analysis hinges upon the common belief that less variability in the service process generally leads to smaller queue-length and delay when the average statistics remains the same.
To this end, we assume for now that the fugacity for each link is a fixed constant. This assumption will be relaxed later in Section~\ref{sec:optimality}, in which we show our algorithm under any finite $T$ is also throughput optimal when the fugacity can be time-varying, set to be some function of the queue-length at each link. At this point, we focus on the long-term behavior of the delay, as any impact of transient phase will eventually fade away as time goes on, and in the standard queueing literature,
the behavior of queue-length or delay is discussed mostly in the steady-state.
We will briefly touch upon the impact of transient phase under our algorithm later in Section~\ref{se:transient}.

First, Little's law asserts that the average delay is determined by the average queue length given that arrival rate is kept the same. For this reason, we are here interested in the stationary behavior of the queue-length driven by the recursion in (\ref{eq:queue}). Let $I(t) = A(t) - \sigma(t)$ be the \emph{net input} to the queue at time $t$, with $\Ex\{I(t)\} = \eta - \pi_{B_v} = -\xi < 0$ for the stability of each queue~\cite{Loynes62}. Let $\bold{A}_t = \sum_{k=1}^t A(k)$ and $\bold{S}_t = \sum_{k=1}^t \sigma(k)$ be the cumulative amounts of arrival and service over $t$ slots, respectively.
For a constant $C$ such that $C>\xi = \pi_{B_v}-\eta> 0$, we define
\begin{equation}
Z(t) \triangleq A(t) - \sigma(t) + C, ~\mbox{and}~ \z_t \triangleq \sum_{k=1}^t Z(k).
\label{def:Z}
\end{equation}
Then, it follows that the recursion in (\ref{eq:queue}) can be written as
\begin{equation}
Q(t) = [Q(t-1) + Z(t)-C]^+,
\label{eq:queue2}
\end{equation}
i.e., the queue-length evolves as if the arrival process is $Z(t)$ with $\Ex\{Z(t)\} = C-\xi$ and the service rate is constant $C$. Thus, the queue-length $Q$ in the steady-state admits the following.
\begin{equation*}
\pr\{Q > x\} = \pr\left\{ \sup_{t \geq 0} \lt[\z_t - Ct\rt] > x \right\}.
\end{equation*}

Our next step is to note that, by similar conditioning on the initial values, we can generalize Propositions~\ref{lm:zero_padding} and \ref{lm:lag_shifting} into the case of multiple random variables, say, $f(X_{t+t_1}), g(X_{t+t_2}),$ $h(X_{t+t_3}),\ldots$ whose time indices are all distinct in modulo $T$. Since the choice of functions is arbitrary, this implies that $T$ distinct sub-processes defined by $\sigma^{(i)}(n) = \{\sigma(nT+i)\}_{n\geq 0}$, $i=1,\ldots,T$ are all independent in the steady state, and the entire correlation structure of the original process $\sigma(n)$ carries over to each of the sub-processes $\sigma^{(i)}(n)$.

\begin{figure}[t!]
    \centering
    \includegraphics[width=3.5in]{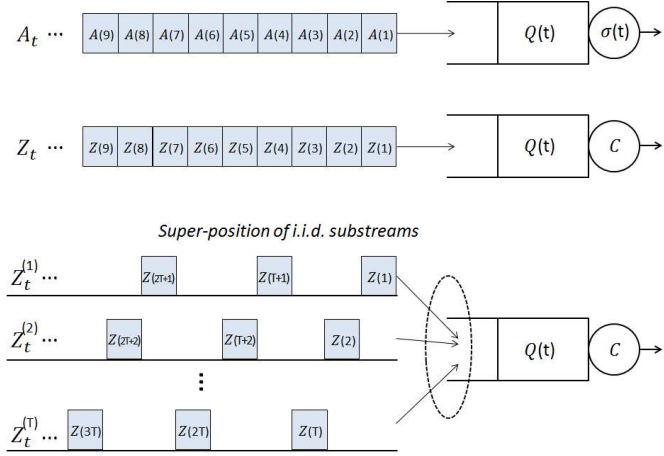}
    \caption{Our algorithm has an effect of dividing the original net-input process of a single stream (top) into a sum of $i.i.d.$ substreams (bottom) in the steady state.} \label{fig:substream}
\end{figure}

In particular, let $Z^{(i)}(n) = Z(nT+i)$, $n\geq 0$, for each $i=1,2,\ldots,T$. From the zero-padding and lag-shifting properties, and since $A(t)$ is $i.i.d.$ over time $t$ and also independent of $\sigma(t)$, it follows that
$\{Z^{(i)}(n)\}_{n\geq 0}$, $i=1,2,\ldots,T$ are $i.i.d.$ processes, each of which has
$\Ex\{Z^{(i)}(n)\} = C-\xi$ and $\cov\{Z^{(i)}(n), Z^{(i)}(n+k)\} = \cov\{Z(n), Z(n+k)\}$ in the steady state. If we consider a cumulative net-input process up to $t=nT$ for some $n\in\mathbb{N}$, then we can decompose $\z_t$ into $\z_t = \sum_{i=1}^T \bold{Z^{(i)}_t}$ where
\begin{align*}
\bold{Z^{(i)}_t}  \triangleq \sum_{k=0}^{n-1} Z^{(i)}(k) & = \bold{A^{(i)}_t} - \bold{S^{(i)}_t} + nC  = \sum_{k=0}^{n-1} \lt[A(kT+i)-\sigma(kT+i)+C \rt].
\end{align*}
Thus, the cumulative `modified net-input' process $\z_t$ to the queue with constant service rate of $C$ is nothing but a superposition of $T$ $i.i.d.$ replicas of the original ($\z_t = \sum_{i=1}^T \bold{Z^{(i)}_t}$), but with its timescale stretched by $T$-fold.
Our algorithm with parameter $T$ thus effectively de-correlates the original service process (or the modified net-input process), regardless of how much correlations persist in the original case caused by any arbitrary topological constraint under the Glauber dynamics. See Figure~\ref{fig:substream} for illustration. Also,
\begin{align*}
\pr\left\{ \sup_{t \geq 0} \lt[\z_t - Ct\rt] > x \right\} = \pr\left\{ \sup_{t \geq 0} \sum_{i=1}^T \lt[\bold{Z^{(i)}_t} - (C/T)t\rt] > x \right\}.
\end{align*}
In other words, the queue at link $v$ behaves as if we aggregate $T$ $i.i.d.$ input and also aggregate $T$ different service capacities of $C/T$ each. Thus, we expect that the usual benefits of statistical multiplexing gain and the economies of scales~\cite{Kelly96,Botvich95a,Hordijk00}, or the principle of ``Sharing resources is always better than partitioning.''~\cite{Kumaran200395}
should apply here.

To quantitatively capture such a gain out of de-correlating the original process under our algorithm, we can also employ the usual Gaussian approximation for $\z_t$, now a sum of $T$ $i.i.d.$ processes, by appealing to the Central Limit Theorem. Note that $\Ex\{\z_t\} = (C-\xi)t$ and we define by $v(t, T)$ the variance of $\z_t$ under our algorithm with order parameter $T$. Then, it is known that the queue-length distribution with Gaussian input can be well approximated by~\cite{Choe00,Eun03,BigQueues}
\begin{equation}
\pr\{Q > x\}
\approx \exp\lt(-\inf_{t>0}\frac{(\xi t+x)^2}{2v(t,T)}\rt)
\label{eq:Gaussian-queue}
\end{equation}
for a wide range of $x>0$. We then have the following.

\begin{proposition} \label{lem:var_T}
Suppose the correlations of $\sigma(t)$ under the original CSMA Glauber dynamics are all non-negative, i.e., $\psi(k) \geq 0$ for all $k \in \mathbb{N}$. Then, for any given $t>0$, $v(t,T)$ is decreasing (non-increasing) in $T>0$.
\end{proposition}

\vspace{2mm} \textbf{Proof:}
Let $\ssigma(t,T)$ be the configuration state generated from our algorithm, a Markov chain of order $T$, and set $\sigma(t, T) =  \idc\{\ssigma(t,T) \in B_v\}$ to be the service process of link $v$ under our algorithm with $T$. Since $\bold{A}_t$ is independent of the service process in any case, and doesn't depend on $T$, it suffices to show that $\var\{\sum_{k=0}^{t-1} \sigma(k,T)\}$ is non-increasing in $T$ for any given $t>0$.

Define $r(k,T) = \cov\{\sigma(0,T), \sigma(k,T)\}$ to be the covariance function in the steady state. We then have
\begin{equation*}
\var\lt\{\sum_{k=0}^{t-1} \sigma(k,T)\rt\} = t\cdot\var\{\sigma(0,T)\} + \sum_{k=1}^{t-1}(t-k) r(k,T).
\end{equation*}
Since the marginal distribution remains the same under our algorithm, $\var\{\sigma(0,T)\}$ does not depend on $T$. Thus, it remains to show that the second term on the RHS is non-increasing in $T$.
Notice that from Propositions~\ref{lm:zero_padding} and \ref{lm:lag_shifting}, and by setting $f(\cdot)=g(\cdot)=\idc\{\cdot \in B_v\}$, we have
\begin{align}
r(k,T) &= 0, ~~~~\text{for}~ k \neq nT, ~n=1,2,\ldots
\label{eq:pro1} \\
r(nT,T) &= r(n,1) \triangleq r(n), \label{eq:pro2}
\end{align}
under our algorithm with parameter $T$.
Now, observe that
\begin{align*}
& \sum_{k=1}^{t-1} (t-k)r(k,T) \stackrel{(\ref{eq:pro1})}{=} \sum_{k=T,2T,\ldots}^{t-1} (t-k)r(k,T) = \sum_{j=1}^{\lfloor \!(t\!-\!1)/T \rfloor} (t-jT)r(jT,T) \stackrel{(\ref{eq:pro2})}{=} \sum_{j=1}^{\lfloor \!(t\!-\!1)/T \!\rfloor} (t-jT) r(j) \\
& \geq \sum_{j=1}^{\lfloor \!(t\!-\!1)/(T\!+\!1)\! \rfloor} \!\!\!(t\!-\!j(T\!+\!1)) r(j) \stackrel{(\ref{eq:pro2})}{=} \sum_{j=1}^{\lfloor \!(t\!-\!1)/(T\!+\!1)\! \rfloor} \!\!\!(t\!-\!j(T\!+\!1)) r(j(T\!+\!1),(T\!+\!1)) \\
& = \sum_{k=T\!+\!1,2(T\!+\!1),\ldots}^{t-1}\!\!\!\!\!\!(t\!-\!k) r(k,(T\!+\!1)) \stackrel{(\ref{eq:pro1})}{=} \sum_{k=1}^{t-1}(t\!-\!k) r(k,(T\!+\!1))
\end{align*}
where the inequality holds since $r(k) = \psi(k)\var\{\sigma(k,1)\} \geq 0$ from the assumption. This completes the proof.
\hfill $\Box$
\vspace{3mm}

We expect that positive temporal correlations in the link service process under the original Glauber dynamics prevail for most cases of network topologies and are in accordance with wide-spread link starvation problem in CSMA scheduling. Recall that we have shown in Proposition~\ref{lem:spectral} that $\psi(k)$ is lower bounded by a well defined positive function for every even $k$. (See also the discussion at the end of Section~\ref{subsection:correlation} for more accounts.) If we consider a lazy chain for the standard Glauber dynamics from the beginning, $\psi(k)\geq 0$ for all $k$ trivially holds. Even in this case, note that our high-order chain approach is guaranteed to reduce the correlations via zero-padding and lag-shifting, leading to the reduction in the variance of the cumulative net-input to the queue, which subsequently results in smaller queue-length via (\ref{eq:Gaussian-queue}).

\subsection{Throughput Optimality} \label{sec:optimality}

In our delayed CSMA algorithm, we set $\lambda_v(t) = e^{W_v(t)}$ for some suitable weight function of link $v$, as discussed in Section~\ref{subse:TO-prelim}. Although the link schedules under our algorithm are updated based on their $T$-step-back states, we set the weight function $W_v(t)$ to be a function of the current queue length $Q_v(t)$ at time $t$, rather than $T$ steps ago, such that the system can react more quickly by adjusting the dynamic fugacities with the latest information.

With the time-varying fugacities, the state transition matrix becomes time-inhomogeneous, which we write as $P_t$, a function of $\lambda_v(t), v \in \N$ at time $t$. For each given such $P_t$, let $\pi_t$ be its unique stationary distribution (in a row vector form), i.e., $\pi_t = \pi_t P_t$, and $\mu_t$ be the actual distribution of the link schedules $\ssigma(t)$ at time $t$ under our algorithm with order parameter $T$. In other words,
we have
\begin{equation}
\mu_t = \mu_{t-T}P_{t-1}.
\label{ttt1}
\end{equation}
Similar to the steps via `network adiabatic' theorem in \cite{SRS09,Shah12-AAP}, a key step in proving the throughput optimality of our algorithm is to show that $\mu_t \approx \pi_t$ for sufficiently large queue lengths under (\ref{ttt1}). The distance between two probability distributions is characterized by the notion of total variation (TV) distance~\cite{Bremaud99,Levin09}, defined as \begin{equation*}
\| \nu - \mu \|_{\text{TV}} = \frac{1}{2} \sum_{\sigma \in \Omega} |\nu(\sigma) - \mu(\sigma)|.
\end{equation*}

Our approach toward the throughput optimality of our proposed algorithm is similar to those in~\cite{SRS09,JRJ10CDC-techrep,Shah12-AAP}, with a little modifications, as follows. First, suppose $P_t$ is changing very slowly over time $t$ for all large queue-lengths. (In fact, this can be achieved by setting the weight function $W_v(t)$ as a very slowly increasing function of the queue-length~\cite{SRS09,JRJ10CDC-techrep,Shah12-AAP}.) Then, the resulting $\pi_t$, a solution to $\pi = \pi P_t$ is also slowly varying over time $t$ such that the actual distribution $\mu_t$ is able to get closer to $\pi_t$ (in the sense of TV distance) before $\pi_t$ moves away to another, thus effectively simulating the separation of timescales. As will be shown in Section~\ref{se:transient}, under our algorithm with order parameter $T$, the speed of convergence of $\mu_t$ under static fugacity (i.e., static $P$) is roughly $T$ times slower than that of the conventional Glauber dynamics. This implies that the speed of actual distribution $\mu_t$ under dynamic fugacity as a slowly varying function of queue-length, will also be reduced by a factor of $T$. Since $T$ is finite, we expect that $\mu_t$ is still able to catch up the slowly varying target $\pi_t$ in time, and by following similar steps in~\cite{SRS09,JRJ10CDC-techrep,Shah12-AAP} our algorithm can also achieve the throughput optimality.
Our next result below shows that this is indeed the case.

\begin{proposition}\label{prop:TO}
Let $\epsilon>0$ be arbitrarily given. For any arrival rate $\eeta \in (1-\epsilon) \mathbb{C}$, set the dynamic link weight\footnote{$Q_{\text{max}}(t)$ is the length of the largest queue in the network at time $t$. In \cite{SRS09} it is argued that $Q_{\text{max}}(t)$ can be easily estimated via any gossip-like message passing mechanism.
The use of $Q_{\text{max}}(t)$ is for purely technical reasons, and it doesn't need to be known in practice, as conjectured in \cite{SRS09,Shah12-AAP}. For simplicity, we assume $Q_{\text{max}}(t)$ is known throughout, and do not discuss this issue here.}
\begin{equation}
W_v(t) = \max \left\{ h(Q_v(t)), \frac{\epsilon}{2|\N|}h(Q_{\text{max}}(t)) \right\},
\label{eq:W_v}
\end{equation}
where $h(\cdot) = \log\log( \cdot + e)$. Then, our algorithm with any finite order parameter $T$ satisfies (\ref{eq:stability}), and is thus throughput optimal.
\end{proposition}

\vspace{2mm} \textbf{Proof:}
See Appendix.
\hfill $\Box$
\vspace{3mm}

Proposition~\ref{prop:TO} together with our delay analysis in Section~\ref{subse:delay} assert that our algorithm with order $T$ achieves not only the required throughput optimality, but also much better delay performance by effectively cutting down the dependency among consecutive link states, thus promoting much faster link state changes, while keeping all the marginal distributions the same.

\section{Impact on Transient Behavior} \label{se:transient}

So far, our focus has been on the study of correlations and throughput-delay performance in the steady-state. As seen in  Section~\ref{se:analysis}, our algorithm is guaranteed to be throughput optimal and yield smaller delay once the system has converged to its stationary regime, and larger $T$ in our algorithm leads to better queueing performance in the steady-state, via correlation reductions in the service process induced by our delayed CSMA (Glauber dynamics of order $T$). While such a convergence itself in a finite-sized system is always guaranteed, it is unclear how our algorithm affects such transient behavior and the speed of convergence to its stationary regime.

To capture such transient behavior,
we utilize a notion of \emph{mixing time} of a Markov chain $P$ with its stationary distribution $\pi$, defined as follows~\cite{Bremaud99,Levin09,Stroock05}
\begin{equation*}
t_{\text{mix}}(\epsilon) \triangleq \min\{t : \max_{x \in \Omega}\|P^{(n)}(x,\cdot) - \pi\|_{\text{TV}} \leq \epsilon, \;\; \forall n \geq t\},
\end{equation*}
where $P^{(n)}(x,\cdot)$ is $n$-step transition probability distribution starting from $x$. For general, non-Markov processes with the same stationary distribution $\pi$ on the same state space, the mixing time can be similarly defined by considering $\|\mu_t - \pi\|_{\text{TV}}$, where $\mu_t$ is the distribution at time $t$ and maximizing over all possible initial distribution $\mu_0$.
Clearly, our delayed CSMA algorithm with order parameter $T$ yields $\mu_t = \mu_{t-T}P$ when the transition kernel is time-homogeneous. Iteratively applying such relation gives $\mu_{nT} = \mu_0 P^{(n)}$, implying that in terms of the mixing time we need roughly $T$ times larger number of state transitions to achieve the same degree of accuracy in distributional error, compared to the conventional CSMA-based algorithm. In other words, in our high-order Markov chain based approach, the stationary distribution of the schedule itself remains untouched, while its induced temporal correlation structure in the steady state is `reshaped' to our advantage, via zero-padding and shifting, to produce the same throughput optimality with better delay performance, all in the same distributed setting as in the usual CSMA. The only caveat is, it takes about $T$ times longer to get there.\footnote{This also suggests that the mixing time based delay analysis may give looser bound on the average queue length \emph{in the steady-state}. See~\cite{SubramanianISIT11} for similar accounts on the usage of mixing time for delay analysis.} This tradeoff between a bit slower convergence but to a `better' stationary regime, also implies that the performance during the transient phase can be worse than that of the conventional CSMA algorithm (albeit our algorithm will eventually pay off in the end), and necessitates careful choice of $T$ depending upon how long the system is meant to run and measured.

\begin{figure}[t!]
    \centering
    \vspace{-0mm}
    \hspace{-10mm}\includegraphics[width=3.3in]{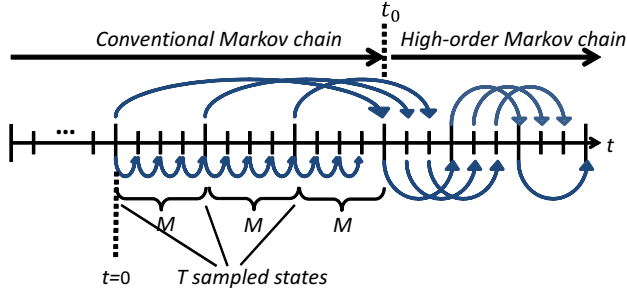}
    \hspace{-14mm}
    \vspace{-3mm}
    \caption{An illustration of the gentler start-up algorithm. Blue arrows indicate state transitions. The algorithm starts at $t_0$ with $T$ initially sampled states.} \label{fig:gentle}
    \vspace{-2mm}
\end{figure}

As a remedy for such a problem, we here provide a gentler start-up algorithm, which combines the strength of relatively faster mixing of the traditional CSMA, and the reduced correlations of our delayed CSMA. This method consists of two steps. First, we run the conventional CSMA algorithm with $T=1$ (so that the mixing time doesn't get hurt) until it gets reasonably closer to its stationary regime, and then \emph{samples} $T$ schedules with inter-spacing $M$ between two consecutive samples, which will then be used as initial states to run our delayed CSMA algorithm. Specifically, after running the traditional CSMA for a while, at a certain time instant agreed among links ($t=0$), each link keeps its channel state information in its memory at every $M$ time slots. It will have $T$ sampled channel states stored after $MT$ time slots. Then, each link performs the delayed CSMA with the sampled states as the first $T$ initial states. See Figure~\ref{fig:gentle} for illustration.

To better understand the effect of $M$, consider a configuration state $X_t$ at time $t$, and suppose the conventional CSMA has run sufficiently long enough so that the marginal distribution can be assumed to be $\pi$, i.e., $\pr\{X_t = x\} = \pi(x)$ for all $t\geq 0$. Without loss of generality, we index time $t=0$ at which the sampling starts. To proceed, we find very useful the following `self-adjoint' property of the operator $\p$ for a reversible chain with respect to $\pi$~\cite{Bremaud99,Levin09,Stroock05}.
\begin{align*}
\langle f,\p g\rangle_{\pi} & = \sum_{x,y \in \Omega}f(x)P(x,y)g(y)\pi(x) \\
& = \sum_{x,y \in \Omega}f(x)P(y,x)g(y)\pi(y) = \langle \p f, g\rangle_{\pi}
\end{align*}
where the second equality is from the reversibility. Iteratively applying the above relation gives
\begin{equation}
\langle\bold{P}^{(n)}f, \bold{P}^{(m)}g\rangle_\pi = \langle\bold{P}^{(n+m)}f, g\rangle_\pi = \langle f, \bold{P}^{(n+m)}g\rangle_\pi,
\label{eq:self-adjoint}
\end{equation}
for any $n,m\geq 0$.

Now, consider the correlation between $X_{t_0+T}$ and $X_{t_0+T+1}$ for example, where $t_0 = MT$ is the time that the delayed CSMA starts as in Figure~\ref{fig:gentle}. Note that
\begin{equation*}
\Ex\{f(X_{t_0})g(X_{t_0\!+\!1})\} = \sum_{x\in \Omega}\Ex\{f(X_{t_0})g(X_{t_0\!+\!1}) | X_{0} \!=\! x\} \pi(x).
\end{equation*}
Conditioning on $X_0 = x$ gives two disjoint sample paths leading to $X_{t_0+T}$ and $X_{t_0+T+1}$ as follows:
\begin{align*}
(i) ~ X_0 \!=\! x & \to X_{t_0} \to X_{t_0+T} \\
(ii) ~X_0 \!=\! x & \to X_{1} \to X_{2} \to \cdots \to X_{M} \to X_{t_0+1} \to X_{t_0+T+1}.
\end{align*}
Thus, by the conditional independence, we have
\begin{align*}
\Ex\{f(X_{t_0\!+\!T})g(X_{t_0\!+\!T\!+1}) ~|~ X_0\!=\!x\} &= \Ex\{f(X_{t_0\!+\!T}) | X_0\!=\!x\} \Ex\{g(X_{t_0\!+\!T+\!1})|X_0 \!=\!x\} \\
& = (\p^{(2)} f)(x) (\p^{(M\!+\!2)} g)(x),
\end{align*}
and therefore
\begin{align*}
&\Ex\{f(X_{t_0\!+\!T}) g(X_{t_0\!+\!T\!+\!1})\} = \sum_{x\in \Omega}(\p^{(2)}f)(x)(\p^{(M\!+\!2)}g)(x) \pi(x) = \langle\p^{(2)}f, \p^{(M\!+\!2)}g \rangle_{\pi} = \langle f, \p^{(M\!+\!4)}g \rangle_{\pi},
\end{align*}
where the last equality is from (\ref{eq:self-adjoint}). This implies that although the time difference between $X_{t_0+T}$ and $X_{t_0+T+1}$ is only one, its correlation behavior is of lag $M\!+\!4$.

For general cases of two points $t_0+i$ and $t_0+j$ with $t_0 = MT$ as in Figure~\ref{fig:gentle}, by counting the number of transitions from $X_0$ as we did in the above, we arrive to the following.
\begin{proposition}\label{prop:corr-reduction}
Let $t_0+i = nT+m$, $t_0+j = n'T+m'$ for $m,m' = 0,\ldots,T\!-\!1$, and $n,n'=1,2,\ldots$. Then,
\begin{equation*}
\Ex\{f(X_{t_0+i})g(X_{t_0+j})\} = \langle f, \p^{(k)}g\rangle_{\pi},
\end{equation*}
where
\begin{align*}
k =
\begin{cases}
  |n-n'|, &\mbox{if}~ m = m', \\
  n+n'+|m-m'|M+2 &\mbox{otherwise.}
\end{cases}
\end{align*}
\end{proposition}

Proposition~\ref{prop:corr-reduction} implies zero-padding and lag-shifting properties as shown in Propositions~\ref{lm:zero_padding} and \ref{lm:lag_shifting}, by taking limits on $i, j$ together and considering $i \not\equiv j \pmod{T}$ and $i \equiv j \pmod{T}$, respectively.\footnote{This is because $\lim_{k \to \infty} r(k) = 0$ since $\pr\{X_k \in B | X_0 \in B\} - \pr\{X_0 \in B\} = \sum_{j=2}^{|\Omega|} \alpha_j \rho^k_j$
from (\ref{eq:sinclair}) and $|\rho_j| < 1$ for $j,=2,\ldots,|\Omega|$.}
The choice of parameter $M$ offers another tradeoff between quicker start-up and convergence speed. For instance, if $M$ is very large so that the first $T$ initial samples are almost independent of each other, then we have zero-padding correlation behavior from the beginning, but it would take very long to prepare such $T$ independent samples. On the other hand, if $M$ is small (say, $M=1$), we have the opposite situation with slower convergence from a quicker start.

\section{Simulation Results}

\begin{figure}[t!]
    \centering
    \includegraphics[width=2.7in,height=1.7in]{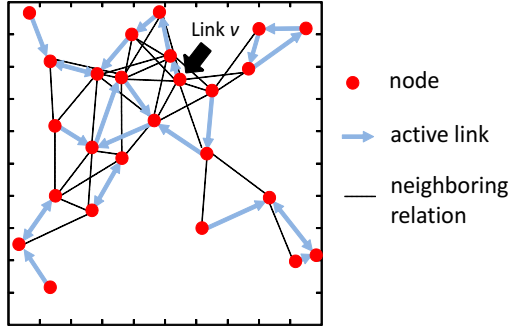}
    \vspace{-2mm}
    \caption{An instance of network configuration with 25 nodes}
    \label{fig:rand_topology}
    \vspace{-1mm}
\end{figure}

In this section, we present simulation results for the delayed CSMA algorithm. We consider a network scenario by a random geometric graph (RGG) model with 25 nodes uniformly and independently positioned over the $1000\times 1000 m^2$ area. The transmission range of each node is set to be $250m$, where two nodes can communicate with each other if they are in the range of transmission. We have each node select one of its neighboring node uniformly at random, and create a communication link. If the receiver node of a link is within a range of transmitter node of another link, we consider that the two links form an edge in the conflict graph. For the decision schedule, we choose the access probabilities $a_i = 0.25$ for all links $i \in \N$. In this scenario, we collect simulation results for the link $v$ shown in Figure~\ref{fig:rand_topology}. In obtaining simulation data, we have taken average of the data from 10000 time repeated simulations, and unless otherwise noted, for each of the runs we discard the data form the first half of the simulation time, to obtain the results in the steady-state.

\begin{figure}[t!]
    \centering
    \vspace{-0mm}
    \hspace{-0mm}\subfigure{\includegraphics[width=3in,height=2.3in]{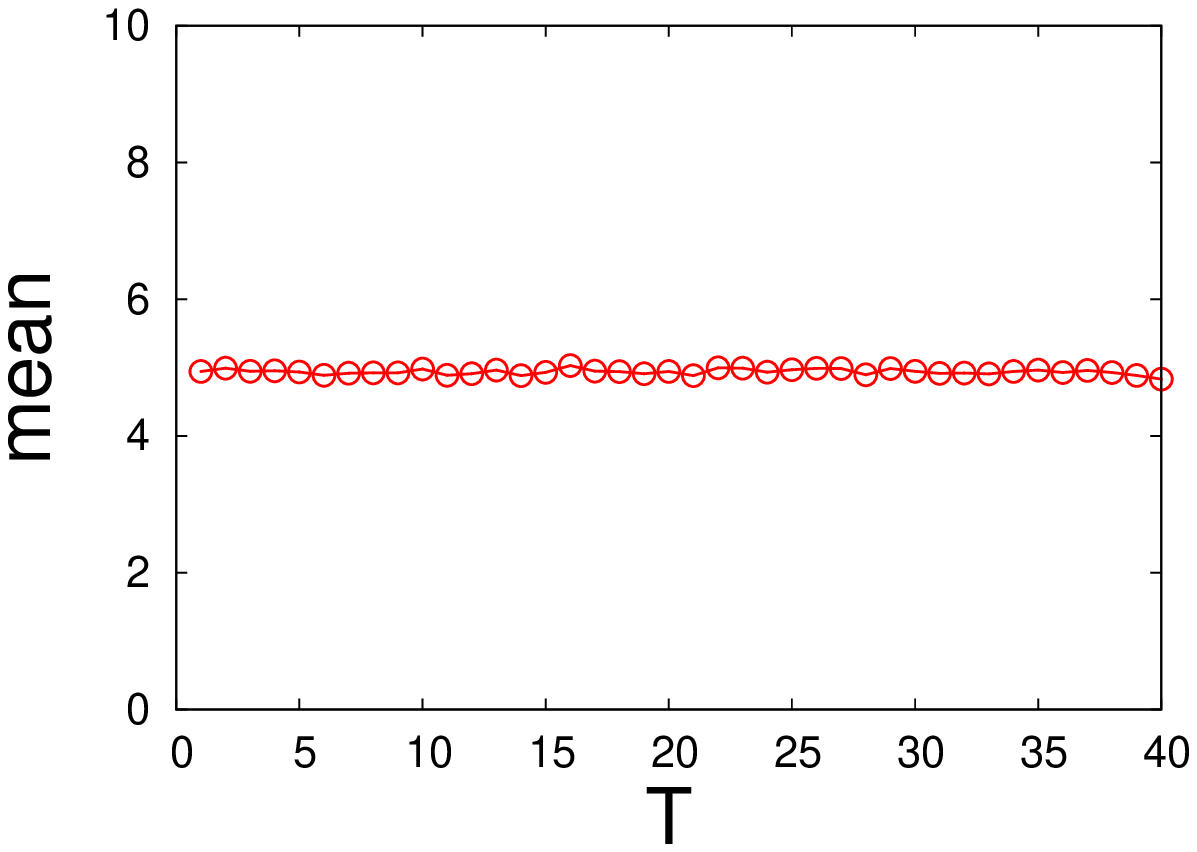}}
    \hspace{-0mm}\subfigure{\includegraphics[width=3in,height=2.3in]{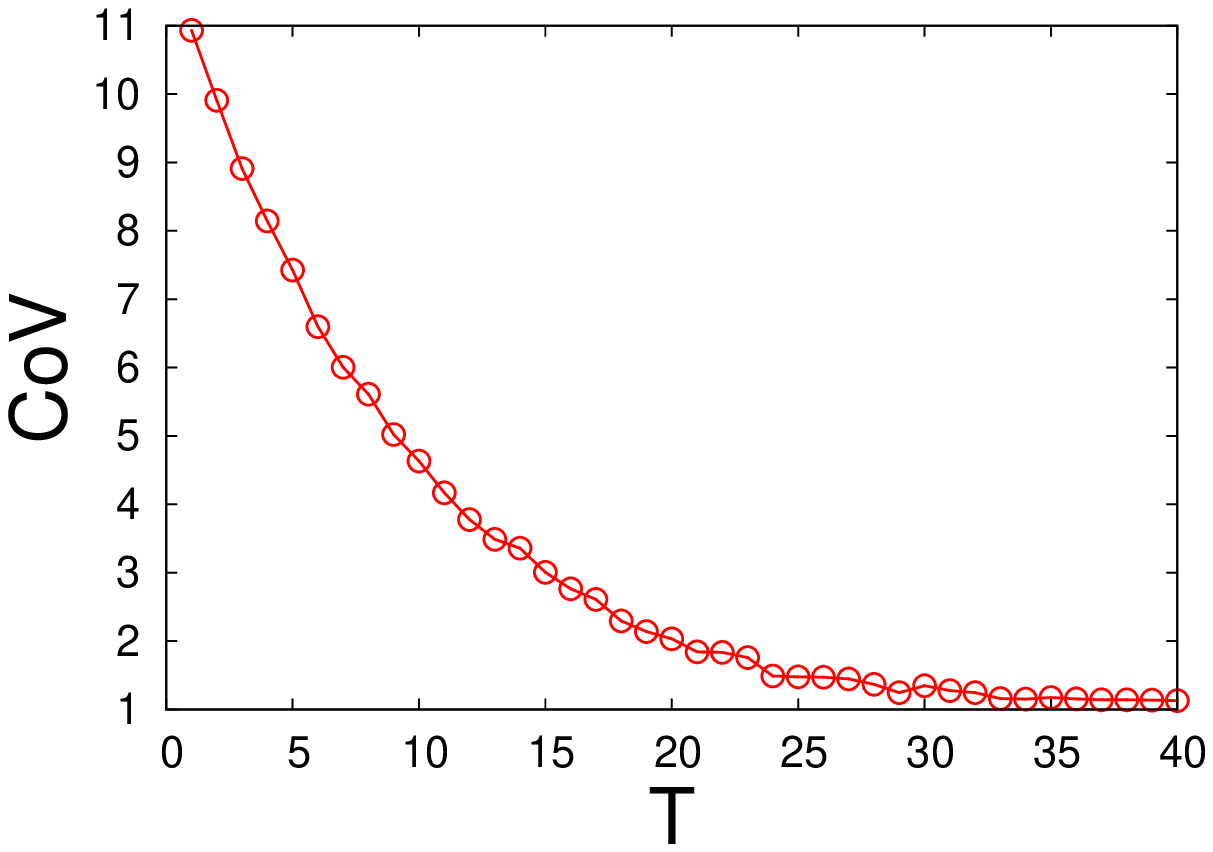}}
    \hspace{-0mm}
    \vspace{-0mm}
    \caption{Mean and CoV of `off' duration for $\sigma_v(t)$} \label{fig:mc}
    \vspace{-0mm}
\end{figure}

First, to understand the benefits of having more drastic changes in the zero-one service process $\sigma_v(t)$ under our delayed CSMA algorithm, we look at the distribution of its `off' duration $U$, the duration from an active slot to the next active slot. We measured its mean $\Ex\{U\}$ and the coefficient of variation (CoV) $\sqrt{\var\{U\}}/\Ex\{U\}$ as we increase the order parameter $T$, where we set $\lambda_i = 1$ for all $i \in \N$. Figure~\ref{fig:mc} shows that the first-order statistics of the link state doesn't change with $T$, while its variability decreases for larger $T$, implying that our algorithm with larger $T$ effectively removes link starvation phenomenon.

\begin{figure}[t!]
    \centering
    \vspace{-0mm}
    \hspace{-0mm}\subfigure[T=1, t=217]{\includegraphics[width=3in,height=2.3in]{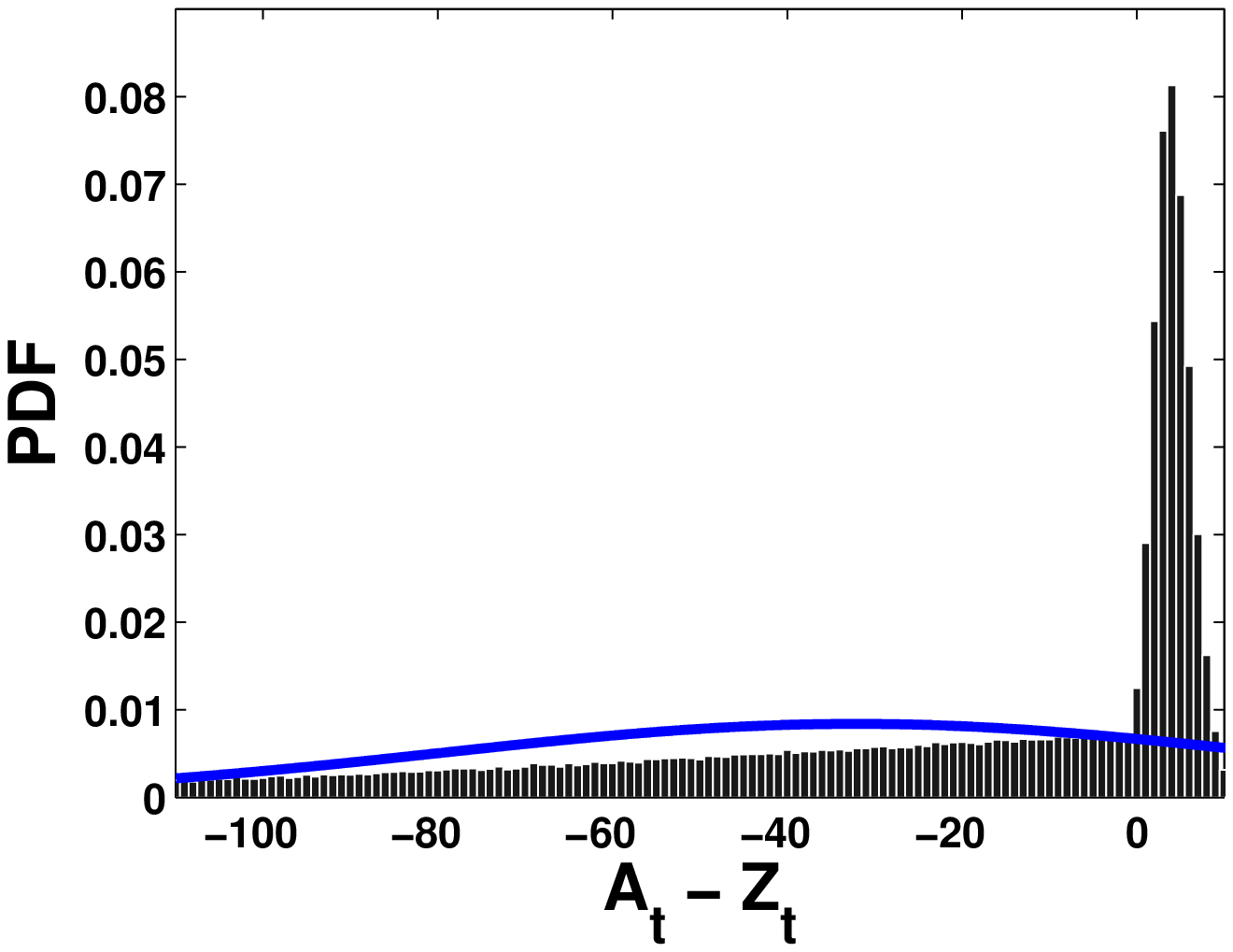} \label{fig:gs1}}
    \hspace{-0mm}\subfigure[T=25, t=217]{\includegraphics[width=3in,height=2.3in]{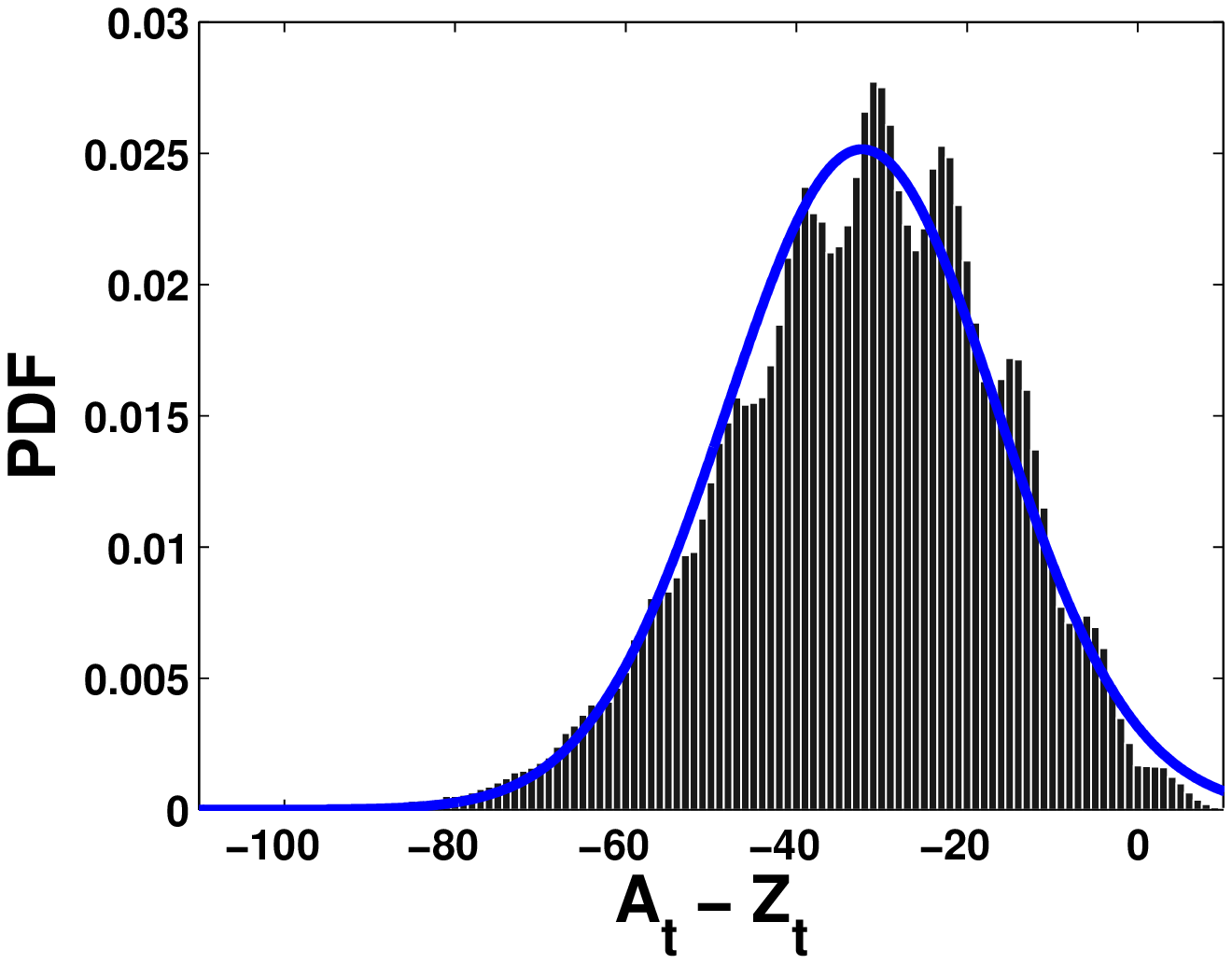} \label{fig:gs2}}
    \hspace{-0mm}
    \vspace{-0mm}
    \caption{Histogram of $\bold{A}_t - \bold{S}_t$ and its approximation by a Gaussian process. The lines are Gaussian distribution drawn from measured sample mean and variance.} \label{fig:gaussian}
    \vspace{-0mm}
\end{figure}

Before investigating the delay performance of our proposed algorithm, we first validate our approximation. As before, under the static fugacity setup, we first collect the cumulative net-input $\bold{A}_t - \bold{S}_t$ up to $t=217$ from a stationary start $t=0$ where its arrival rate is set to be $\eta_v = 0.1$.
As seen in Figure~\ref{fig:gs1}, the net-input process in the standard CSMA case ($T=1$) is far from being Gaussian due to the strong correlation structure in $\sigma_v(t)$. As we increase the order parameter to $T=25$, it becomes very close to Gaussian as shown in Figure~\ref{fig:gs2}. The same conclusion also applies to the modified net-input $\z(t) = \bold{A}_t - \bold{S}_t + Ct$.

\begin{figure}[t]
    \centering
    \vspace{-0mm}
    \hspace{-0mm}\subfigure{\includegraphics[width=3in,height=2.3in]{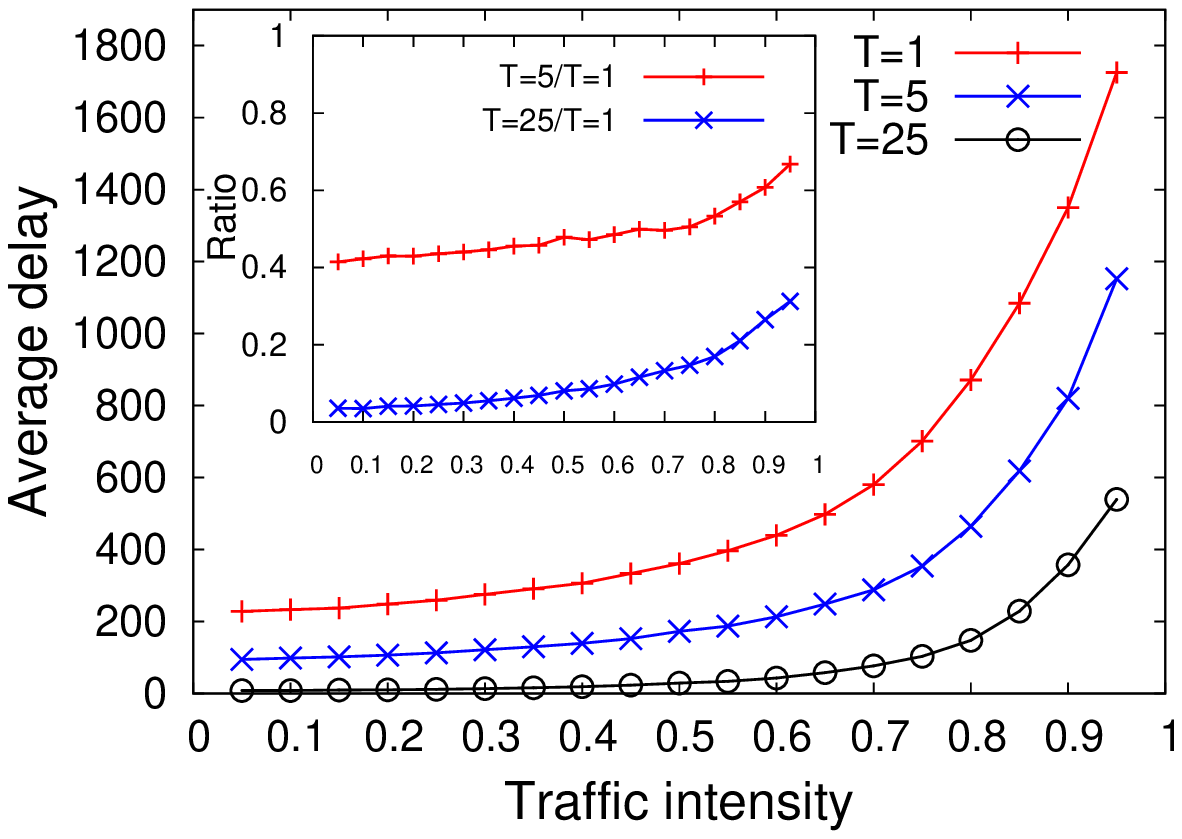}}
    \hspace{-0mm}\subfigure{\includegraphics[width=3in,height=2.3in]{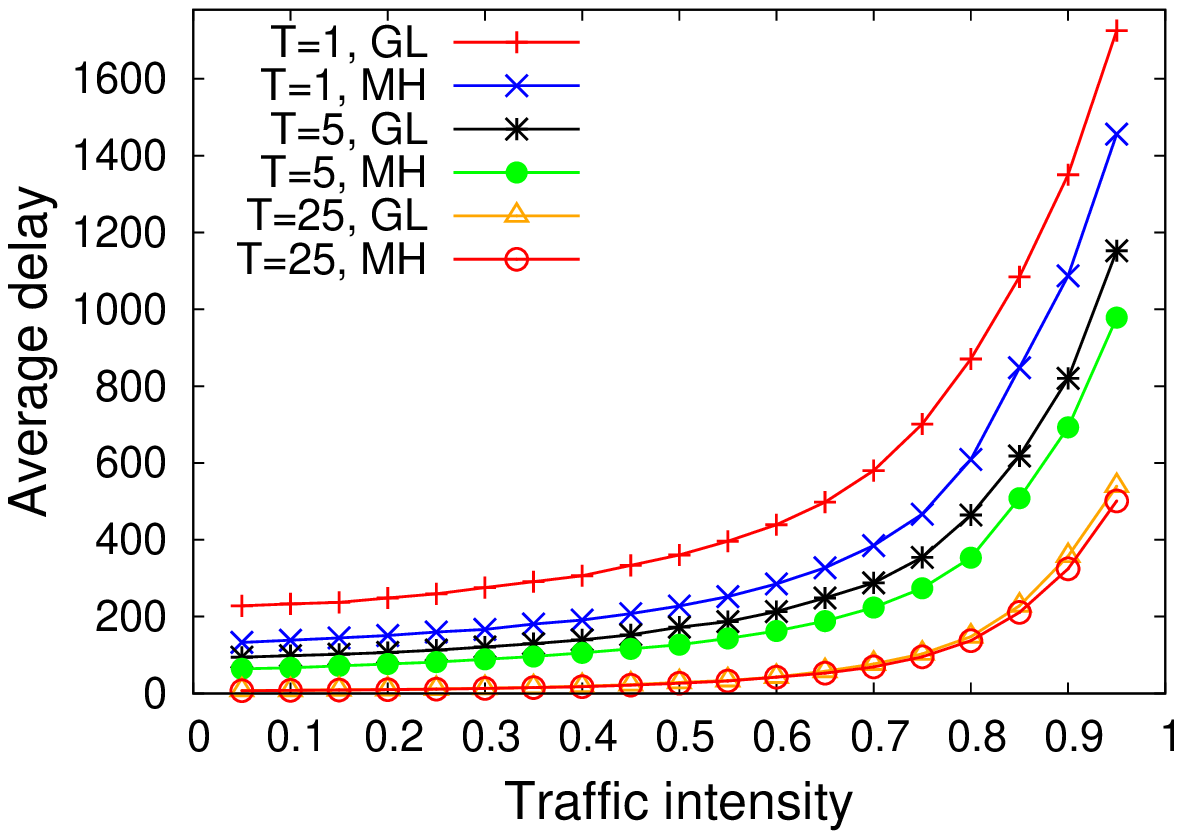}}
    \hspace{-0mm}
    \vspace{-0mm}
    \caption{Delay performance of the delayed CSMA algorithm with various $T$ under dynamic fugacity}
    \label{fig:delay}
    \vspace{-0mm}
\end{figure}

In measuring delay performance, we consider dynamic fugacity with the weight function set to be $\log\log(Q_i(t)+e)$ for all $i \in \N$. We have measured the delay of each packet in the queue until it gets served, under different traffic intensity. We adjusted arrival rate of each link by gradually increasing the rate proportional to one of its potential link capacities, which is calculated by summing over all possible maximal independent sets with equal weight. As the traffic intensity increases to 1, the rate approaches to maximum throughput.
The left-hand side of Figure~\ref{fig:delay} shows the delay improvement over the conventional CSMA algorithm, where the inset figure displays the ratio of the delay under our algorithm with chosen $T$ to that of standard CSMA. We note that the performance improvement is quite remarkable. For instance, with $T=5$, the delay reduces by half, and with $T=25$, it reduces by a factor of 20 compared to the conventional CSMA algorithm. We observed similar trends in the improvement under different weigh functions such as $\log(Q_i(t)+1)$ or $Q_i(t)$.

\def\subfigcapskip{-2pt}

Recall that our approach is to add `order of $T$' into the original Glauber dynamics to keep the same marginal distribution while shaping the correlation structure to our advantage. We note that this approach will hold for any other standard algorithm modeled by a reversible Markov chain on the same feasible state space achieving the same stationary distribution, not necessarily the Glauber dynamics considered so far. For example, it was shown in~\cite{Clee12} that Metropolis-Hastings (MH) based algorithm (still reversible Markov chain on $\Omega$) outperforms the conventional Glauber dynamics. The right-hand side of Figure~\ref{fig:delay} shows that indeed MH algorithm gives better delay performance than the standard Glauber dynamics, but still, the amount of delay reduction is significant for larger $T$. Interestingly, the performance gap between the standard Glauber and the MH algorithm (all with $T=1$) seems to narrow down for larger $T$, implying that the effect of correlation reductions via high-order Markov chain dominates over the choice of base reversible Markov chains.

\begin{figure}[t!]
    \centering
    \vspace{-0mm}
    \hspace{-10mm}\subfigure[Traffic intensity=0.1]{\includegraphics[width=3in,height=2.3in]{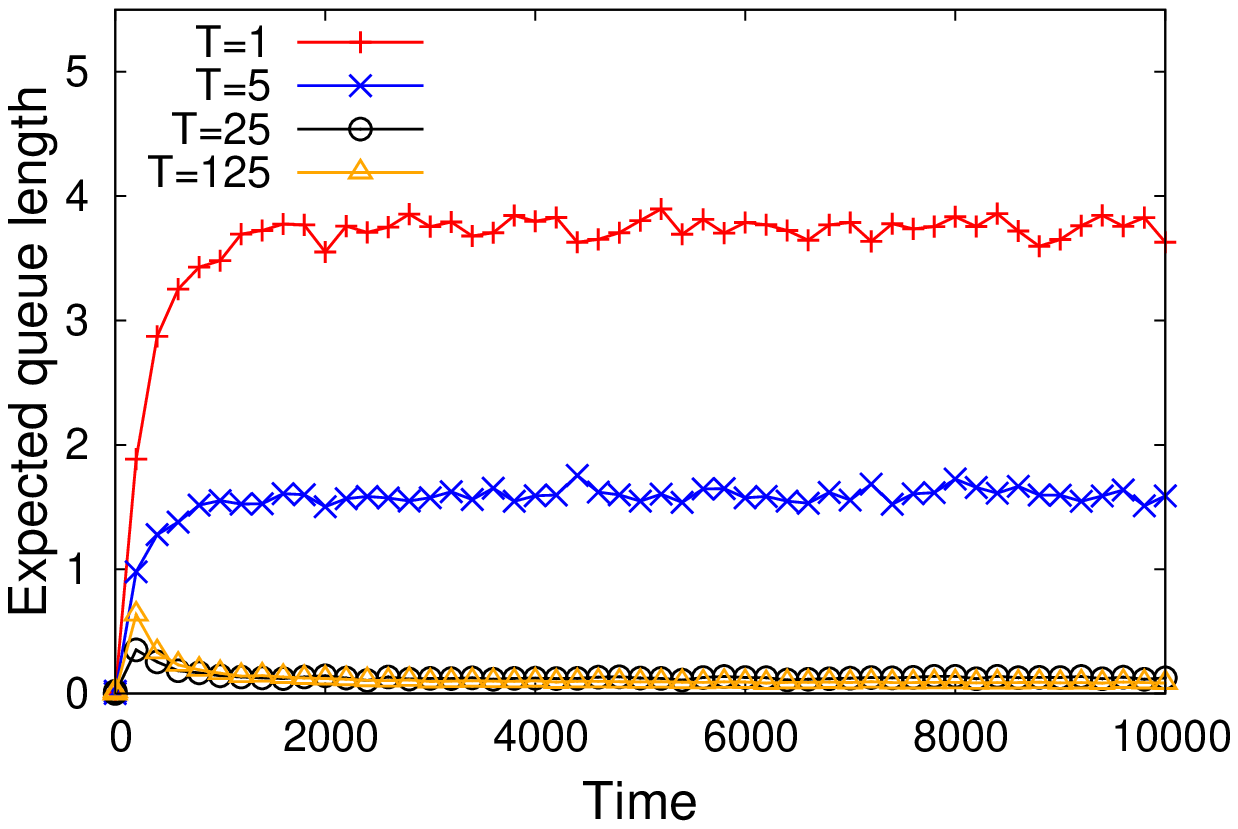} \label{fig:tr1}}
    \hspace{-4mm}\subfigure[Traffic intensity=0.9]{\includegraphics[width=3in,height=2.3in]{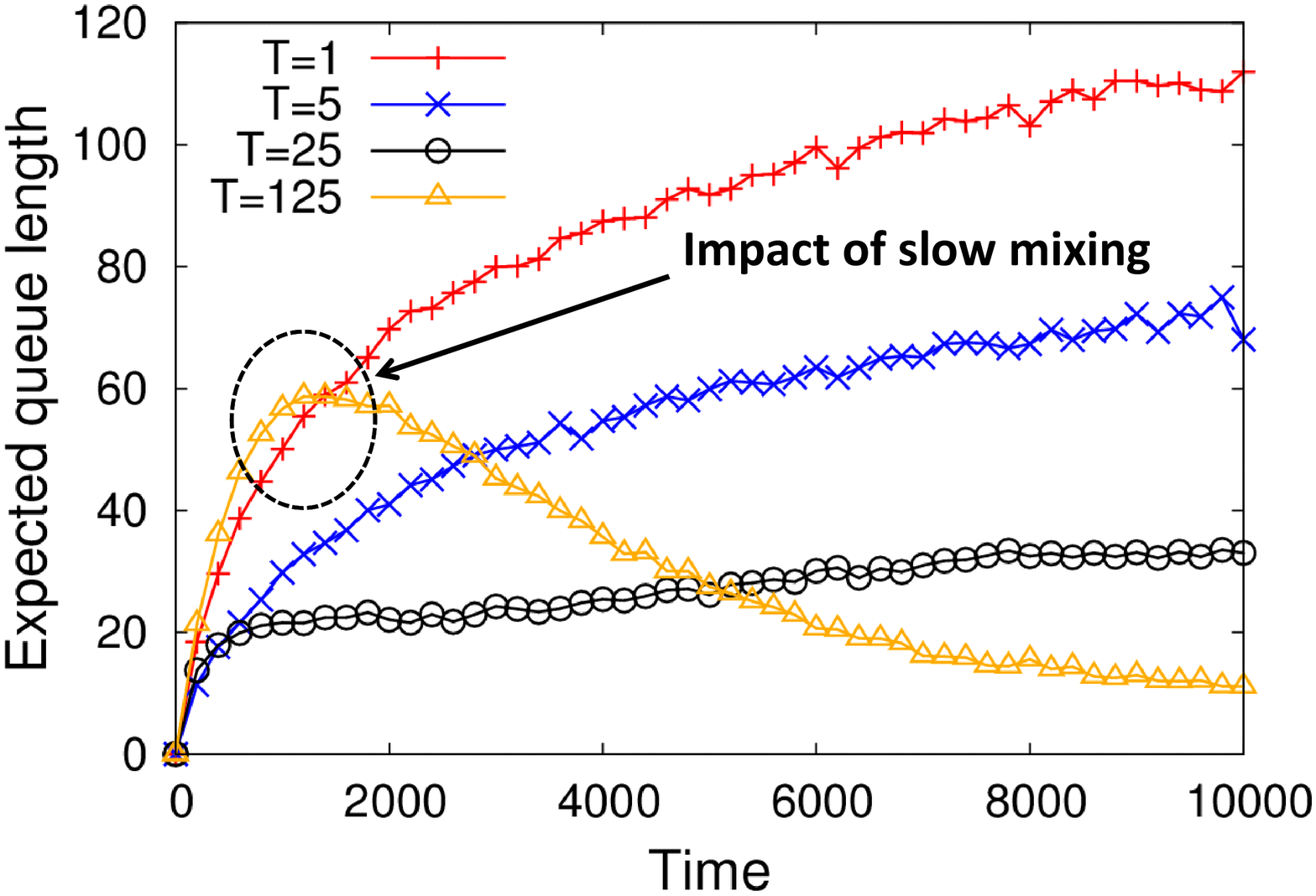} \label{fig:tr2}}
    \hspace{-14mm}
    \vspace{-2mm}
    \caption{Impact of $T$ on queue evolution.} \label{fig:transient_q}
    \vspace{-2mm}
\end{figure}

Now we look at the impact of transient behavior induced by our algorithm with different $T$. We have measured the queue-length from after simulation starts. In order to measure the queue-length during the initial phase, we collect queue-length samples and obtain its time average for different time indexes. Figure~\ref{fig:transient_q} shows the measured queue-length under different $T$.
As discussed in section \ref{se:transient}, we might expect larger backlog at initial phase when large $T$ is used (due to slower mixing time). However, in practice, its impact is not significant unless indefinitely large $T$ is used. For example, as can be seen from Figure~\ref{fig:tr1}, its impact is almost negligible when traffic intensity is low. Even under relatively high traffic intensity (Figure~\ref{fig:tr2}) and with large $T$, e.g., $T=125$, the accumulation of backlog at transient phase due to slower mixing is still comparable to that of the original case ($T=1$), and the queue size eventually converges to smaller one in the steady state. Thus, we expect that the benefit in long-term behavior will outweigh such a drawback in most of situations in practice.

\begin{figure}[t!]
    \centering
    \vspace{-0mm}
    \hspace{-0mm}\subfigure[$\Ex\{Q(t)\}$ with $T=50$]{\includegraphics[width=3in,height=2.3in]{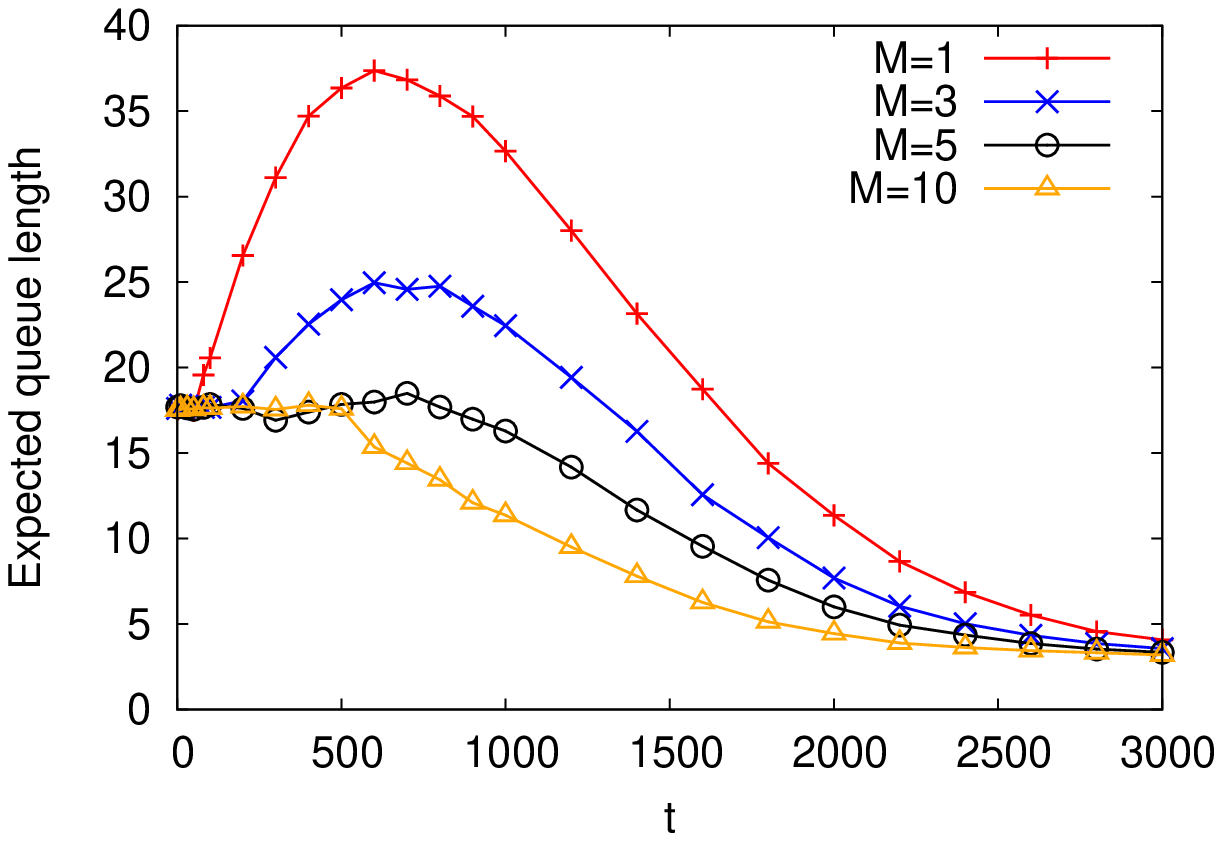}\label{fig:Ma}}
    \hspace{-0mm}\subfigure[$\psi(k)$ with $T=5$]{\includegraphics[width=3in,height=2.3in]{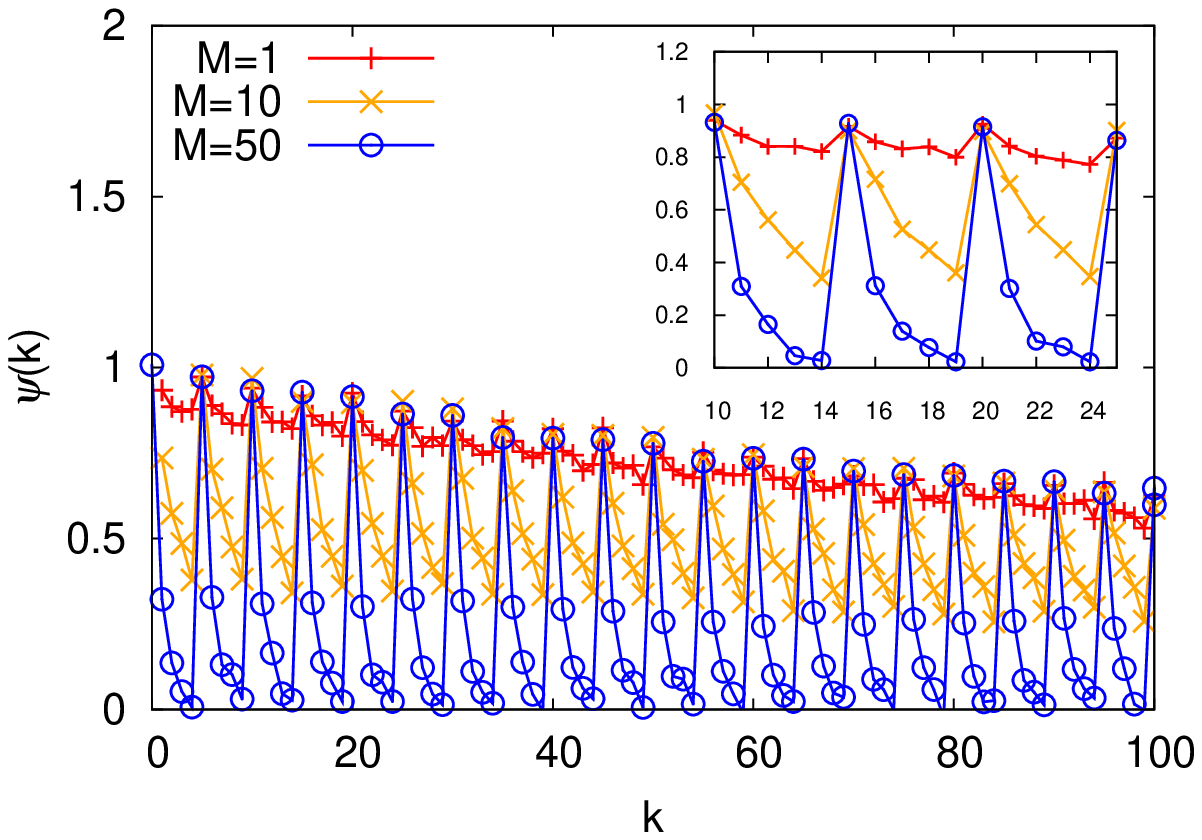}\label{fig:Mb}}
    \hspace{-0mm}
    \vspace{-0mm}
    \caption{Impact of $M$ on (a) queue-length evolution and (b) correlation reduction}
    \label{fig:effect_M}
    \vspace{-0mm}
\end{figure}

We also look at the benefit of utilizing another parameter $M$; the gentler start scheme proposed in section \ref{se:transient}. In this case, we have run simulations under static fugacity with $\lambda_i = 1$ for all links $i \in \N$ in order to better observe the de-correlation process induced by the parameter $M$. Firstly, we have run the conventional CSMA algorithm for sufficiently long enough time, and then measured the queue length right after the delayed CSMA algorithm starts to run. Figure~\ref{fig:Ma} shows the measured queue-length with $T=50$ under different $M$. For $M=1$ (quick start), large order parameter $T=50$ leads to worse performance than the standard CSMA at initial phase, but its strong correlation reduction property in the end eventually offsets such a drawback. As discussed in Section\ref{se:transient}, such an effect in the transient phase can be alleviated by sampling intermediate states that are separated by $M$. We can see that fairly small values of $M$ can readily reduce the drawback. Larger $M$ leads to quicker correlation reduction as seen in Figure~\ref{fig:Mb} (for lags not a multiple of $T=5$ here), and also as predicted from Proposition~\ref{prop:corr-reduction}.

\section{Conclusion}

In this paper, we have proposed the delayed CSMA algorithm based on a high-order Markov chain on the same state space of feasible schedules achieving the same stationary distribution, but with completely different and also reshaped correlation structure. Our algorithm is extremely simple to implement without requiring any additional message passing or overhead, with the exception of a single parameter $T$ across the network once and for all. We have proved that our algorithm is throughput-optimal, yet provides far better delay performance by emulating the effect of superposition of independent traffic streams in the queue, out of a single input and still single service process, via zero-padding and lag-shifting properties of the resulting service process of the queue under our algorithm.

Another interesting viewpoint out of our investigation is that, the conventional approach via mixing time for delay performance must be used with great care, since in our setting we achieve better queueing performance in the steady-state by trading much less correlations for a bit slower mixing speed. While fast-mixing and less correlations typically come in pair under the traditional Markov chain based approach, we note that our high-order Markov chain (or simply non-Markov on the same state space) can now separate these two, thus enabling us to trade one for the other, toward better understanding and design of distributed schedulers running over different timescales of interest.

\section{Acknowledgement}

This material is based upon work supported by the National Science Foundation under Grant Number \#1217341. Any opinions, findings, and conclusions or recommendations expressed in this material are those of the author(s) and do not necessarily reflect the views of the National Science Foundation.

\bibliographystyle{abbrv}
\bibliography{ref-all,myref}

\newpage\newpage
\normalsize
\newpage
\allowdisplaybreaks

\begin{center}
\textbf{\LARGE Appendix}
\end{center}

\textbf{Proof of Proposition~\ref{prop:TO}:}
First, we need the following.
\begin{lemma} \cite{ESP05} \label{lemma:optimality}
For a scheduling algorithm, given any $0< \epsilon, \delta < 1$ and
if there exists $0 < B(\delta, \epsilon) < \infty$ such that in any time slot $t$, with probability larger than $1-\delta$, the scheduling algorithm chooses a schedule $\sigma(t) \in \Omega$ that satisfies
\begin{equation}
\sum_{v \in \ssigma(t)} W_v(t) \geq (1- \epsilon)\max_{\ssigma \in \Omega} \sum_{v \in \ssigma} W_v(t) \label{eq:epsilon_maxweight}
\end{equation}
whenever $\max_{v \in \N} Q_v(t) > B(\delta, \epsilon)$, then the scheduling algorithm is throughput-optimal in the sense of (\ref{eq:stability}).
\end{lemma}

We basically extend the procedure used for proving the throughput optimality for conventional CSMA algorithm in~\cite{JRJ10CDC-techrep}. The key steps therein can be summarized as follows. (See \cite{JRJ10CDC-techrep} for details.)
\begin{enumerate}
\item Given $\delta \in (0,1)$, find a $K(\delta)$ such that
\begin{equation}
\| \pi_t - \mu_t \|_{TV} \leq \frac{\delta}{4}
\end{equation}
holds whenever $Q_{\max}(t) \geq K(\delta)$.
\item Given $\epsilon \in (0,1)$, choose $B(\delta, \epsilon)$ as
\begin{equation}
B(\delta, \epsilon) = \text{max}\left\{ K(\delta), h^{-1}\left(\frac{|\N|\log2 + \log{\frac{2}{\delta}}}{\epsilon/2} \right) \right\}.\nonumber
\end{equation}
\item Then for any arrival rate $\eeta \in (1-\epsilon) \mathbb{C}$, Lemma~\ref{lemma:optimality} can be established and thus the scheduling algorithm is throughput optimal.
\end{enumerate}

Since the delayed CSMA algorithm yields $\mu_t = \mu_{t-T}P_{t-1}$ as opposed to $\mu_t = \mu_{t-1}P_{t-1}$ in the standard CSMA algorithm, we will have to prove the step 1 above in our setting, and finding a $K(\delta)$ under our algorithm with order parameter $T$ is the key challenge in proving throughput optimality. This statement is established in the following.

\begin{lemma} \label{lemma_main}
Given any $\delta\in (0,1)$ and under our delayed CSMA with parameter $T$,
there exists $K(\delta)<\infty$ such that $\| \mu_t(\sigma) - \pi_t(\sigma) \|_{\text{TV}} \leq \frac{\delta}{4}$ holds whenever $Q_{\max}(t)\geq K(\delta)$.
\end{lemma}

The rest of this paper is devoted to the proof of Lemma~\ref{lemma_main} and this will complete our proof of Proposition~\ref{prop:TO}.
To proceed, we collect some notations and useful lemmas first.
\begin{definition} \cite{Bremaud99}
(\emph{$\chi^2$ distance}) The $\chi^2$ distance between two probability distributions $\nu$ and $\mu$ on a finite space $\Omega$ is defined by
\begin{equation*}
\big\|\nu - \mu \big\|_{\frac{1}{\mu}}^2 = \sum_{\sigma \in \Omega} \frac{1}{\mu(\sigma)} \left(\nu(\sigma) - \mu(\sigma) \right)^{2}.
\end{equation*}
\end{definition}
Then, the following relationship holds \cite{Bremaud99}.
\begin{equation}
\big\| \nu - \mu \big\|_{\frac{1}{\mu}} = \lt\| \frac{\nu}{\mu} - 1 \rt\|_{\mu} \geq 2 \| \nu - \mu \|_{TV}. \label{eq:tv_bound}
\end{equation}

\begin{lemma}
(\emph{The $1/\pi$-bound}) \cite{Bremaud99} Let $P$ be an aperiodic, irreducible, and reversible transition matrix on the finite space $\Omega$, with its stationary distribution $\pi$. Let $1 = \rho_1 > \rho_2 \geq \cdots \geq \rho_{|\Omega|} > -1$ be the eigenvalues of $P$.
Then for any probability distribution $\nu$ on $\Omega$, and for all $n \geq 1$,
\begin{equation}
\big\| \nu P^n - \pi \big\|_{\frac{1}{\pi}} \leq \rho(P)^n \big\| \nu - \pi \big\|_{\frac{1}{\pi}} \label{eq:pi_bound}
\end{equation}
where $\rho(P) = \max\{\rho_2, |\rho_{|\Omega|} | \}$ is the second largest eigenvalue modulus (SLEM) of $P$.
\end{lemma}

\begin{lemma} \label{lemma:multiple_lag}
  Given $t, k \in \mathbb{N}$, define
  \begin{equation}
  \alpha_t = \sum_{i \in \N} \left\{h'(\hat{Q}_v(t)) + h'(\hat{Q}_v(t+1)) \right\} \nonumber
  \end{equation}
  where $\hat{Q}_v(t) = h^{-1}(W_v(t))$. If $\alpha_t < \frac{1}{k}$, then
  \begin{enumerate}
  \item $1-k\alpha_t \leq \frac{\pi_{t+k}(\sigma)}{\pi_{t}(\sigma)} \leq 1+k\alpha_t$, $\forall \sigma \in \Omega$.
  \item $\big\| \pi_{t+k} - \pi_{t} \big\|_{\frac{1}{\pi_{t+k}}} \leq 2k \alpha_t$.
  \end{enumerate}
\end{lemma}
The above is a generalized version of Lemma 13 in \cite{SRS09}, and is straightforward to verify, so we omit the proof here.

In determining the decision schedule $m(t)$ at time $t$, without loss of generality, we assume a single site update rule, i.e., only one link is selected uniformly at random at each time slot. Our analysis can then be extended to the case of multiple site updates by following the same steps of Lemma 7 in~\cite{JRJ10CDC-techrep}. Thus, we will mainly refer to the analysis for single site update as given in Lemma 3 of~\cite{JRJ10CDC-techrep}, which is reproduced below. From this point on, for notational simplicity, we set $N$ as the number of links in the conflict graph, i.e., $N = |\N|$.

\begin{lemma} Let $M_t = \frac{1}{1-\rho(P_t)}$, where $\rho(P_t)$ is the SLEM of transition probability matrix $P_t$. Then, we have
\begin{equation}
M_{t} \leq 16^N e^{(4N W_{\max}(t))} \label{eq:mixing_time}
\end{equation}
where $W_{\max}(t) = \max_{v \in \N} W_v(t)$
\end{lemma}

We provide the following lemma that is essential to our proof, which is a modification of Lemma 14 in \cite{SRS09}, but with our order parameter $T$ in mind.
\begin{lemma}
  Given any $\delta\in (0,1)$, define a constant $B = B(N,\delta)$ satisfying $B \!\geq\! (16N\!-\!1)^{16N\!-\!1}$ and
  \begin{equation}
  \frac{64T16^NN \log^{4N}(x\!+\!T\!-\!1\!+\!e)}{ \exp\lt((\log(x+e))^{\frac{\delta}{2N}}\rt) -1-e} < \delta \label{eq:B}
  \end{equation}
  for all $x \geq B$. If $Q_{\max}(t) \geq B$, then
  \begin{equation}
  M_{t+T-1}\cdot \alpha_t \leq \frac{\delta}{32T}, ~~\mbox{and}~~ M_{t+T-1}\cdot \alpha_{t-1} \leq \frac{\delta}{32T}. \nonumber
  \end{equation}
\end{lemma}

\vspace{2mm} \textbf{Proof:}
Note that $\hat{Q}_{\text{min}} = h^{-1}(W_{\text{min}}) \geq h^{-1}\left(\frac{\delta}{2N} h(Q_{\text{max}})\right)$ from (\ref{eq:W_v}) and $h^{-1}$ is increasing. Using $h'(x) = \frac{1}{(x+e)\log(x+e)} < \frac{1}{x}$ for $x>0$, and the fact that $Q(t)-k \leq Q(t+k) \leq Q(t)+k$ for any $k\geq 1$, we have, for $i\!=\!0,1$,
\begin{equation}
\alpha_{t-i} \leq \frac{N}{\hat{Q}_{min}(t\!-\!i)} + \frac{N}{\hat{Q}_{min}(t\!+\!1\!-\!i)} \leq \frac{2N}{\hat{Q}_{min}(t)\!-\!1} \nonumber
\end{equation}
Also, from (\ref{eq:mixing_time}), $M_{t\!+\!T\!-\!1} \leq 16^N \log^{4N}(\hat{Q}_{\max}(t)\!+\!T\!-\!1\!+\!e)$.
Then, for $i = 0, 1$,
\begin{align}
M_{t\!+\!T\!-\!1} \cdot \alpha_{t-i} &\leq \frac{2N16^N \log^{4n} \left(\hat{Q}_{\max}(t)\!+\!T\!-\!1\!+\!e\right)}{\hat{Q}_{min}(t)-1}  \nonumber \\
&\leq \frac{2N16^N \log^{4n}(x\!+\!T\!-\!1\!+\!e)}{\small e^{(\log(x+e))^{\frac{\delta}{2N}}}\!-\!1\!-e} \label{eq:ratio_bound}
\end{align}
where $x := \hat{Q}_{\max}(t) \geq B$. From (\ref{eq:B}) and
since the term in (\ref{eq:ratio_bound}) goes to zero as $x\to\infty$, it is bounded above by $\frac{\delta}{32T}$.
\hfill $\Box$
\vspace{3mm}

The following is the main statement in proving the throughput optimality, which is also similarly used in \cite{SRS09,JRJ10CDC-techrep}.
\begin{lemma}\label{lemma:main-TO2}
At time $t$, if $Q_{\max}(t) \geq B+t^*$ where  $t^*$ is
\small
\begin{equation*}
T^2 \left\lceil 16^{2N}\log^{8N}(B\!+\!T\!+\!e) \log\left(\frac{4}{\delta}(2\log(B\!+\!T\!-\!1\!+\!e))^{N/2}\right) \right\rceil^2
\end{equation*}
\normalsize
Then,
\begin{equation}
\big\| \mu_t - \pi_t \big\|_{\frac{1}{\pi_t}} \leq \delta/2. \label{eq:adiabatic}
\end{equation}
\end{lemma}

\vspace{2mm} \textbf{Proof:}
We follow similar steps as in the proof of Lemma~12 in~\cite{SRS09} with some modifications. First, by applying triangle inequality, we have
\begin{align}
\big\| \mu_t - \pi_t \big\|_{\frac{1}{\pi_t}} &\leq \big\| \mu_t - \pi_{t-1} \big\|_{\frac{1}{\pi_t}} + \big\| \pi_{t-1} - \pi_{t} \big\|_{\frac{1}{\pi_t}} \nonumber \\
&\leq (1+\alpha_{t-1}) \big\| \mu_t - \pi_{t-1} \big\|_{\frac{1}{\pi_{t-1}}} + 2\alpha_{t-1}, \nonumber
\end{align}
where the second inequality is from Lemma \ref{lemma:multiple_lag}, and the fact that $\alpha_{t-1} < \frac{\delta}{32} < 1$.
We can see that (\ref{eq:adiabatic}) holds if
\begin{equation}
\big\| \mu_t - \pi_{t-1} \big\|_{\frac{1}{\pi_{t-1}}} \leq \delta/4. \nonumber
\end{equation}

Define $r_j = \| \mu_{j} - \pi_{j-1} \|_{\frac{1}{\pi_{j-1}}}$. Using the Lemma (\ref{eq:pi_bound}) and triangle inequality again, we have
\begin{align}
r_{j+T} &= \big\| \mu_{j+T} \!-\! \pi_{j\!+\!T\!-\!1} \big\|_{\frac{1}{\pi_{j\!+\!T\!-\!1}}} \nonumber \\
&= \big\| \mu_{j} P_{j\!+\!T\!-\!1} \!-\! \pi_{j\!+\!T\!-\!1} \big\|_{\frac{1}{\pi_{j\!+\!T\!-\!1}}} ~~~\mbox{(from (\ref{ttt1}))}
 \nonumber \\
&\leq \rho(P_{j\!+\!T\!-\!1}) \big\| \mu_{j} \!-\! \pi_{j\!+\!T\!-\!1} \big\|_{\frac{1}{\pi_{j\!+\!T\!-\!1}}} \nonumber \\
&\leq \rho(P_{j\!+\!T\!-\!1}) \left( \| \mu_{j} \!-\! \pi_{j} \|_{\frac{1}{\pi_{j\!+\!T\!-\!1}}} \!+\! \| \pi_{j} \!-\! \pi_{j\!+\!T\!-\!1} \|_{\frac{1}{\pi_{j\!+\!T\!-\!1}}}\! \right). \label{eq:transition}
\end{align}
Let $t_0$ be the latest time that $Q_{\max}$ hits $B$, i.e., $t_0 = t - \Delta t_0$, where
\begin{align*}
\Delta t_0 = \text{min}\{\Delta t \geq 0 : Q_{\max}(t-\Delta t) = B\}.
\end{align*}
Since $Q_{\max}(j) \geq B$ for all $t_0 \leq j \leq t$, we have for such $j$,
\begin{align}
\| \mu_{j} - \pi_{j} \|_{\frac{1}{\pi_{j\!+\!T\!-\!1}}} & \leq (1+(T-1)\alpha_{j}) \| \mu_{j} - \pi_{j} \|_{\frac{1}{\pi_{j}}} \nonumber \\
&\leq (1+(T-1)\alpha_{j}) ((1+\alpha_{j-1}) r_j + 2 \alpha_{j-1}) \nonumber \\
&\leq \left(1\!+\!\frac{\delta}{32M_{j\!+\!T\!-\!1}} \right) \left( \left(1\!+\!\frac{\delta}{32M_{j\!+\!T\!-\!1}}\right)r_j \! + \! \frac{\delta}{16M_{j\!+\!T\!-\!1}}\right) \label{eq:rt1}
\end{align}
where the last inequality is from $(T\!-\!1)\alpha_j \leq \frac{(T\!-\!1)\delta}{32TM_{j\!+\!T\!-\!1}} \leq \frac{\delta}{32M_{j\!+\!T\!-\!1}}$.
Similarly,
\begin{equation}
\big\| \pi_{j} - \pi_{j\!+\!T\!-\!1} \big\|_{\frac{1}{\pi_{j\!+\!T\!-\!1}}} \leq 2(T\!-\!1)\alpha_{j} \leq \frac{\delta}{16M_{j\!+\!T\!-\!1}}. \label{eq:rt2}
\end{equation}
If we assume $r_{j} < \delta/4$, then it can be checked that (\ref{eq:rt1}) + (\ref{eq:rt2}) $< \frac{\delta}{4} + \frac{\delta}{4M_{j\!+\!T\!-\!1}}$, and hence from (\ref{eq:transition}), (\ref{eq:rt1}), and (\ref{eq:rt2}),
\begin{align*}
r_{j+T} & < \rho(P_{j\!+\!T\!-\!1}) \left(\frac{\delta}{4} + \frac{\delta}{4M_{j\!+\!T\!-\!1}}\right) = \left( 1- \frac{1}{M_{j\!+\!T\!-\!1}} \right) \left(\frac{\delta}{4} + \frac{\delta}{4M_{j\!+\!T\!-\!1}}\right) \leq \frac{\delta}{4}.
\end{align*}
Therefore, if there exists some $\tilde{k}$ that satisfies the conditions \textbf{(C1)} $r_{t-\tilde{k}T}\leq \delta/4$, and \textbf{(C2)} $\tilde{k}\leq\frac{t-t_0}{T}$, then we will have $r_t \leq \delta/4$.

Let $k'$ be the largest $k$ such that $t_0 \leq t - kT$, and let $t'_0 = t - k'T$. Then it can be verified that finding $\tilde{k}$ is equivalent to finding $k^*$ such that \textbf{(C1')} $r_{t'_0+k^*T} \leq \delta/4$, and \textbf{(C2')} $k^*\leq\frac{t-t_0}{T}$. To find $k^*$, assume $r_{t'_0+kT} > \delta/4$ for all $k<k^*$, then we get the following upper bound for $j = t'_0+T, t'_0+2T,\ldots,t'_0+k^*T$.
\small
\begin{align*}
r_j &\leq \left(\!1\!-\!\frac{1}{M_{j\!-\!1}}\!\right) \! \left(\! \left(\!1\!+\!\frac{\delta}{32M_{j\!-\!1}}\!\right)^2\! r_{j\!-\!T}\! +\! \frac{\delta}{16M_{j\!-\!1}}\!\left(1\!+\!\frac{\delta}{32M_{j\!-\!1}}\!\right)\! \right) \\
&\leq \left(\!1\!-\!\frac{1}{M_{j\!-\!1}}\!\right) \! \left(\! \left(\!1\!+\!\frac{\delta}{32M_{j\!-\!1}}\!\right)^2\! r_{j\!-\!T}\! +\! \frac{r_{j\!-\!T}}{8M_{j\!-\!1}}\left(\!1\!+\!\frac{\delta}{32M_{j\!-\!1}}\right)\! \right) \\
&\leq \left(\!1\!-\!\frac{1}{M_{j\!-\!1}}\!\right) \! \left(\! 1\!+\!\frac{1}{M_{j\!-\!1}}\! \right)\!r_{j\!-\!T} = \left(\!1\!-\!\frac{1}{M^2_{j\!-\!1}}\right) r_{j\!-\!T} ~\leq~ e^{-\frac{1}{M^2_{j\!-\!1}}} \cdot r_{j\!-\!T}.
\end{align*}
\normalsize
Then, we have $r_{t'_0 + k^*T} \leq r_{t'_0} \exp\left(-\sum_{j=1}^{k^*} \frac{1}{M^2_{t'_0+jT}}\right),$ where
\begin{align}
\sum_{j=1}^{k^*} \frac{1}{M^2_{t'_0+jT}} &\geq \sum_{j=1}^{k^*} \frac{1}{16^{2N}e^{8Nf(Q_{\max}(t'_0+jT))}} \nonumber \\
&= \sum_{j=1}^{k^*} \frac{1}{16^{2N} (\log(Q_{\max}(t'_0+jT)+e))^{8N}} \nonumber \\
&\geq \sum_{j=1}^{k^*} \frac{1}{16^{2N} (\log(Q_{\max}(t'_0)+k^*T+e))^{8N}} \nonumber \\
&= \frac{k^*}{16^{2N} (\log(Q_{\max}(t'_0)+k^*T+e))^{8N}} \nonumber \\
&\geq \frac{\sqrt{k^*}}{16^{2N} (\log(Q_{\max}(t'_0)+1+e))^{8N}\sqrt{T}} \nonumber \\
&\geq \frac{\sqrt{k^*}}{16^{2N} (\log(Q_{\max}(t_0)+T+e))^{8N}\sqrt{T}}. \nonumber
\end{align}
The third inequality is from the fact that $k^*\geq 1$, $Q_{\max}(t_0) = B \geq (16N-1)^{16N-1}$ and the following inequality in \cite{SRS09},
\small
\begin{equation}
\sqrt{x} \geq \left(\frac{\log(x+y)}{\log(1+y)}\right)^{8N}, \;\; \forall x\geq 1, y \geq(16N-1)^{16N-1}. \nonumber
\end{equation}

\normalsize
In addition,
\begin{align*}
r_{t'_0} = \Big\| \mu_{t'_0+1} - \pi_{t'_0} \Big\|_{\frac{1}{\pi_{t'_0}}} \leq \sqrt{\frac{1}{ \min_{\ssigma\in\Omega}\pi_{t'_0}(\ssigma)}} \leq \sqrt{Z(t'_0)} & \leq (2e^{h(Q_{\max}(t'_0))})^{N/2} \\
& = (2\log(Q_{\max}(t_0)\!+\!T\!-\!1\!+\!e))^{N/2},
\end{align*}
where $Z(t)$ is the normalizing constant for the distribution $\pi_t$, i.e.,
$Z(t) = \sum_{\ssigma \in \Omega} \prod_{v \in \N} (\lambda_v(t))^{\sigma_v}$ with $\lambda_v(t) = e^{W_v(t)}$. If we choose $k^*$ as,
\begin{align}
T \left\lceil 16^{2N}\log^{8N}(B\!+\!T\!+\!e) \log\left(\frac{4}{\delta}(2\log(B\!+\!T\!-\!1\!+\!e))^{N/2}\right) \right\rceil^2 \nonumber
\end{align}
\normalsize
then it can be checked that $r_{t'_0+k^*T} \leq \delta/4$, so \textbf{(C1')} is satisfied. And since $Q_{\max}(t_0) = B$ and $Q_{\max}(t) \geq B+t^*$ (note $t^* = k^*T$) as given in the lemma, we can check $t \geq t_0 + t^*$, which implies \textbf{(C2')}. This completes the proof of Lemma~\ref{lemma:main-TO2}.
\hfill $\Box$
\vspace{3mm}

Therefore, the proof of Lemma~\ref{lemma_main} completes by the choice of $K(\delta) = B(N,\delta) + t^*$, and from the inequality in (\ref{eq:tv_bound}).

\end{document}